\documentclass[a4paper,onecolumn,11pt]{quantumarticle}
\pdfoutput=1

\usepackage{amsmath,amsfonts,amssymb,amsthm}
\usepackage{mathtools}
\usepackage{setspace}
\usepackage[normalem]{ulem} 
\usepackage{physics}
\usepackage{verbatim}
\usepackage{xspace}

\usepackage{graphicx,xcolor}
\usepackage{hyperref}
\hypersetup{pdfstartview=}
\usepackage{url}

\usepackage{verbatim}
\usepackage{alltt}
\usepackage{moreverb}
\usepackage{bm}
\usepackage{bbold}

\usepackage{xr}
\usepackage[export]{adjustbox}
\usepackage{tabularray}
\usepackage[percent]{overpic}

\numberwithin{equation}{section}

\theoremstyle{definition}

\newcommand{\bexp}[1]{\exp\left[#1\right]}
\newcommand{\listb}[1]{\{#1\}}
\newcommand{\listp}[1]{\left(#1\right)}
\newcommand{\trb}[2][]{\text{tr}_{#1}\left[#2\right]}
\newcommand{\btr}[2][]{\text{tr}_{#1}\left[#2\right]}

\newcommand{\shot}{trial}
\newcommand{\shots}{trials}
\newcommand{\tin}{\text{in}}
\newcommand{\gerr}{\varepsilon}

\newcommand{\blog}[1]{\log \left[#1\right]}

\newcommand{\prechosenparam}{\theta_{i_0}}
\newcommand{\paramrefpoint}{\boldsymbol\theta^{(0)}}
\newcommand{\fr}{fully-randomized}

\usepackage{multirow}
\usepackage[caption=false,position=top,justification=raggedright,singlelinecheck=false]{subfig}

\providecommand{\ignore}[1]{}

\newif\ifcmnt
\cmnttrue
\ifdefined\cmntsoff\cmntfalse\fi

\ifcmnt

\else

\fi

\usepackage{tikz}
\usetikzlibrary{backgrounds}
\begin{document}
\title{Optimized experiment design and analysis for \fr{} benchmarking}
\author{Alex Kwiatkowski}
\affiliation{National Institute of Standards and Technology, Boulder, Colorado 80305, USA}
\affiliation{Department of Physics, University of Colorado, Boulder, Colorado, 80309, USA}

\author{Laurent J. Stephenson}
\affiliation{National Institute of Standards and Technology, Boulder, Colorado 80305, USA}
\affiliation{Department of Physics, University of Colorado, Boulder, Colorado, 80309, USA}

\author{Hannah M. Knaack}
\affiliation{National Institute of Standards and Technology, Boulder, Colorado 80305, USA}
\affiliation{Department of Physics, University of Colorado, Boulder, Colorado, 80309, USA}

\author{Alejandra L. Collopy}
\affiliation{National Institute of Standards and Technology, Boulder, Colorado 80305, USA}

\author{Christina M. Bowers}
\affiliation{National Institute of Standards and Technology, Boulder, Colorado 80305, USA}
\affiliation{Department of Physics, University of Colorado, Boulder, Colorado, 80309, USA}

\author{Dietrich Leibfried}
\affiliation{National Institute of Standards and Technology, Boulder, Colorado 80305, USA}

\author{Daniel H. Slichter}
\affiliation{National Institute of Standards and Technology, Boulder, Colorado 80305, USA}

\author{Scott Glancy}
\affiliation{National Institute of Standards and Technology, Boulder, Colorado 80305, USA}
\author{Emanuel Knill}
\affiliation{National Institute of Standards and Technology, Boulder, Colorado 80305, USA}
\affiliation{Center for Theory of Quantum Matter, University of Colorado, Boulder, Colorado 80309, USA}

\newcommand{\dataMainBasicOrig}{2.42^{+0.30}_{-0.22} \times 10^{-5}}
\newcommand{\dataMainBasicOrigFR}{2.80^{+0.14}_{-0.14} \times 10^{-5}}
\newcommand{\dataMainBasicOptFR}{2.57^{+0.07}_{-0.06} \times 10^{-5}}
\newcommand{\dataMainMomentsOrig}{2.39^{+0.65}_{-0.43} \times 10^{-5}}
\newcommand{\dataMainMomentsOrigFR}{3.73^{+0.77}_{-0.52} \times 10^{-5}}
\newcommand{\dataMainMomentsOptFR}{2.67^{+0.08}_{-0.08} \times 10^{-5}}

\begin{abstract}

  Randomized benchmarking (RB) is a widely used strategy to assess the quality of available quantum gates in a computational context. 
  RB involves applying known random sequences of gates to an initial state and using a final measurement step to determine ‘success’ or ‘failure’ for each trial. 
  The probabilities of success and failure over many trials can be used to determine an effective depolarizing error per step of the sequence, which is a metric of the gate quality.
  Here
  we investigate the advantages of \fr{} benchmarking,
  where a new random sequence is drawn for each experimental trial. 
  The advantages of full randomization include smaller confidence intervals on the inferred step error, 
  straightforward optimization of the sequence lengths, 
  the ability to model and measure behaviors that go beyond the typical assumption of time-independent error rates,
  and the ability to use maximum-likelihood inference and to perform statistical tests that rely on maximum-likelihood inference. 
  We discuss models of time-dependent or non-Markovian errors that generalize the basic RB model of a single exponential decay of the success probability. For any of these models, we implement a concrete protocol to
  minimize the uncertainty of the estimated parameters with a fixed constraint on
  the time for the complete experiment, and we implement a
  maximum likelihood analysis. Furthermore, we consider several previously published experiments
  and determine the potential for improvements with
  optimized full randomization. We experimentally observe such improvements in Clifford randomized benchmarking experiments on a single trapped ion qubit at the National Institute of Standards and Technology (NIST). 
  For an experiment with uniform lengths and intentionally repeated sequences the step error was $\dataMainBasicOrig$, and for an optimized fully-randomized experiment of the same total duration the step error was $\dataMainBasicOptFR$. We find a substantial decrease in the uncertainty of the step error as a result of optimized fully-randomized benchmarking.
\end{abstract}
\maketitle

\section{Introduction}
Benchmarking quantum gates is an important task for the design, development, and characterization of a quantum processor \cite{eisert_quantum_2020}. 
Randomized benchmarking is a widely-used method to benchmark gates in a computational context that takes advantage of long sequences of gates to efficiently gain statistical information even when large errors in state preparation and measurement (SPAM) are present \cite{PhysRevLett.117.060505,Knill_2008,Erhard_2019,Helsen_2022,wallman_randomized_2018,proctor.PhysRevLett.119.130502,huang_fidelity_2019,PhysRevLett.125.080504,heinrich_general_2023,meier_thesis,Magesan_2012,PhysRevLett.106.180504}.
In general, a randomized benchmarking trial consists of state preparation, followed by a random sequence of steps drawn from a carefully chosen distribution, followed by a measurement indicating  `success' or `failure' for the trial depending on whether or not the nominal outcome is observed.
The simplest example is standard randomized benchmarking, where each step is drawn from a two-design \cite{Dankert_two_design,meier_thesis}.
If each step is modeled as a perfect unitary gate followed by an error channel, and if that error channel is independent of time, sequence position, and the gate that is applied, then the probability of the `success' outcome decays exponentially as a function of sequence length \cite{Magesan_2012}.
The rate of this exponential decay is interpreted as the average fidelity of a step in a computational context.
In addition to standard randomized benchmarking, a wide variety of other randomized benchmarking variants have been proposed and studied in recent years.
For example, some variants are designed to characterize the fidelity of gates that aren't compatible with two-designs \cite{Carignan_Dugas_2015}, and other variants are designed for inference of system parameters other than the average fidelity of steps \cite{PhysRevLett.123.060501, Wallman_2015}.
In many such situations the behavior of the success probability as a function of sequence length can deviate from a single exponential decay \cite{helsen_new_2019, Hashagen2018realrandomized, Carignan_Dugas_2015, PhysRevLett.109.240504}.

Due to hardware limitations, current experimental implementations of all variants of randomized benchmarking intentionally repeat each random sequence many times to collect statistics about the probability of error.
The main hardware limitation for most experiments is the fact that the time required to load a new random sequence into the relevant control systems, which are nearly always based on field programmable gate arrays (FPGAs), is very large compared to the time required to run a sequence that has already been loaded.
Therefore, it has been more efficient for most experiments to run repeated trials of the same random sequence. 
Here we study \fr{} benchmarking, where a new random sequence is drawn for each experimental trial.
As quantum hardware is maturing, it has become easier for experiments to implement \fr{} benchmarking.
One of the main reasons for this is that more of the experimental design and control is being moved to the FPGA level in hardware, thereby avoiding the latency and nondeterministic timing of loading large lists of instructions or control waveform values from a separate computer into the FPGA.
As a demonstration of this possibility, as part of our work we run a \fr{} benchmarking experiment on a single trapped-ion qubit, where the random sequences are generated on the fly by the control FPGA.
Other research groups are currently developing similar capabilities.
As one example, we refer to Ref.~\cite{fruitwala2024hardwareefficientrandomizedcompiling}, which describes a recent experimental implementation of randomized Pauli gates at the FPGA level.
Another such example can be found in Ref.~\cite{caune2024demonstratingrealtimelowlatencyquantum}, which describes real-time decoding of syndrome measurements performed on an FPGA.
Both of these systems could be adapted for \fr{} benchmarking. 
Going forward, we expect that an increasing amount of the classical side-processing that is necessary for quantum computing will be performed at the FPGA level. 
Furthermore, performing randomized benchmarking of fault-tolerantly encoded (``logical'') qubits necessarily requires a substantial amount of fast, low latency classical processing that is run concurrently with the actual quantum experiment. 
Consequently, fully randomizing the benchmarking of logical qubits should incur very little additional overhead, for any number of logical qubits.
In this paper we present a rigorous statistical study of \fr{} benchmarking that will be useful to experimentalists as the community moves toward fully-randomized benchmarking.

When we compare \fr{} benchmarking to randomized benchmarking with intentionally repeated sequences, we find several concrete advantages of \fr{} benchmarking.
Broadly, these advantages come from the fact that, for an arbitrary error channel, a \fr{} benchmarking experiment is statistically indistinguishable from one where the error channel is a depolarizing channel with the same fidelity.
The same is not true if sequences are intentionally repeated. If the sequences are intentionally repeated the true error channel determines the distribution of success probabilities over the possible random sequences, and properties of this distribution are observable in the statistics of repeated sequences. 
We also note that any randomized benchmarking variant, such as those described in Ref. \cite{Helsen_2022}, can be made \fr{}, and in fact most theoretical treatments of randomized benchmarking implicitly or explicitly assume that the experiment is \fr{}. 

A more detailed summary of the advantages of \fr{} benchmarking is as follows.
First, a randomized benchmarking experiment that repeats random sequences generally has a larger uncertainty in the step error when compared to the same experiment where the sequences are \fr{}.
The larger uncertainty comes from increased variance in the estimate of success probability at each sequence length due to the distribution of fidelities over all possible random sequences.
This has been studied in Ref.~\cite{Wallman_2014} where the authors recommend designing experiments with relatively short sequence lengths in order to mitigate the effect of repeating random sequences. 
They also argue that the effect from not fully randomizing is small for Pauli error channels, but point out that their arguments do not apply to unitary error channels.
In fact, we analyze previously published randomized benchmarking experiments and find evidence in some cases that a significantly smaller uncertainty could have been obtained if the experiment were \fr{}.
A second advantage of \fr{} benchmarking is that the choice of the set of sequence lengths and the choice of the number of \shots{} for each sequence length can be optimized in a straightforward way to maximize the information gained during the experiment.
In experiments that do not fully randomize, optimization of the experiment design requires knowledge of least-squares weights that are generally unknown and depend on the true error model. 
Previous work about optimization strategies for randomized benchmarking can be found in Refs.~\cite{PhysRevA.99.052350,granade_accelerated_2015,itoko_rb_opt_2021}.
Ref.~\cite{PhysRevA.99.052350} provides a heuristic optimization strategy for the basic exponential decay model of randomized benchmarking and suggests choosing a short sequence length and a long sequence length at half the inverse step error.
Ref.~\cite{granade_accelerated_2015} addresses optimization for a Bayesian inference procedure, and Ref.~\cite{itoko_rb_opt_2021} discusses optimization strategies for non-fully-randomized benchmarking where a particular variance model is chosen.
A third advantage of \fr{} benchmarking is that the step error can be inferred by using maximum likelihood in a straightforward way. In contrast, in experiments that intentionally repeat sequences the step error is typically inferred by means of a weighted-least-squares fit with weights that are a priori unknown, which complicates the interpretation of confidence intervals.
Finally, \fr{} benchmarking allows for a straightforward analysis of the simplest time-dependent or sequence-position-dependent errors.
In \fr{} benchmarking, the only effect of these errors, or of any non-Markovian errors, is to modify the behavior of the success probability as a function of sequence length.
We introduce a nested sequence of statistical models for \fr{} benchmarking that can be used to detect this modified behavior in a straightforward way.
A key benefit of \fr{} benchmarking is that maximum likelihood inference makes it possible to perform likelihood-ratio tests to compare models' fits to the data.
Furthermore, the reduced uncertainty of an optimized \fr{} experiment allows for the detection of modified behavior with increased statistical significance.

This paper is organized as follows.
In Section~\ref{sec_overview} we provide an overview of randomized benchmarking, including our conventions for depolarizing channels and step errors. 
In Section~\ref{sec_opt} we describe the numerical procedure that we use to optimize the design of an experiment to minimize the uncertainty in the step error according to a pre-chosen statistical model and reference point. 
In Appendix~\ref{sec_opt_le} we explain how this optimization is equivalent to a maximization of Fisher information.
In Section~\ref{sec_other_models} we address other statistical models of \fr{} benchmarking that allow for time-dependent, sequence-position-dependent, or other non-Markovian errors leading to non-exponential decay of the success probability.
In Section~\ref{sec_comparisons} we make comparisons between previously published randomized benchmarking experiments and optimized \fr{} benchmarking experiments and demonstrate that improvements in uncertainty are possible.
In Section~\ref{sec_fr_variance} we analyze the improvement in uncertainty from \fr{} benchmarking in terms of the underlying distribution of success probabilities over random sequences at a fixed sequence length.
We give evidence that the improvement in uncertainty can be significant when the underlying errors are unitary.
In Section~\ref{sec_statistics} we describe the statistical analysis we use to infer the step error, which consists of maximum likelihood inference and statistical bootstrapping to obtain confidence intervals.
We also describe an empirical likelihood ratio test that we use to possibly reject the basic model of a single exponential decay and demonstrate it on simulated data.
In Section~\ref{sec_exp_implementation} we report the results of randomized benchmarking experiments run on a single trapped ion qubit at NIST.
We implement \fr{} benchmarking and perform a comparison between randomized benchmarking with uniformly chosen sequence lengths and repeated sequences, \fr{} benchmarking with uniformly chosen sequence lengths, and optimized \fr{} benchmarking, under otherwise equal conditions. We find that substantial reductions in uncertainty are possible.

\newcommand{\jmax}{j_{\text{max}}}
\section{Overview of randomized benchmarking}
{
\label{sec_overview}
\newcommand{\modelplist}{\boldsymbol{\theta}}
We describe our conventions and notation for a statistical description of \fr{} benchmarking experiments, with a focus on the basic model of a single exponential decay.
For further information and discussion of randomized benchmarking in general, we refer to Refs.~\cite{Magesan_2012,Knill_2008,meier_thesis,Helsen_2022}.
A \fr{} benchmarking experiment consists of many independent \shots{}, where a \shot{} of sequence length $n$ is composed of a state preparation, followed by a random sequence of $n$ steps, followed by measurement indicating `success' or `failure'.
The content and distribution of the random steps depends on the benchmarking variant in use.
For example, in standard randomized benchmarking each step nominally implements a random Clifford gate.
The design of a \fr{} experiment consists of the list $\listp{n_{j}}_{j=1}^{\jmax}$ of sequence lengths to be used,
and the list $\listp{w_{j}}_{j=1}^{\jmax}$ of the numbers of independently-randomized \shots{} to be performed at each sequence length. 
In a \fr{} experiment, the order in which the total $\sum_j w_j$ \shots{} are performed should also be randomized. As we discuss at the end of this section, this can help to minimize the effect of potential time-dependent errors.
After the experiment is performed, the data consists of a list $\listp{c_j}_{j=1}^{\jmax}$ where $c_j$ is number of success counts observed out of $w_j$ total trials at the sequence length $n_j$.

In general, a \fr{} benchmarking experiment is described by a function $P(n)$ for the average success probability over all sequences of length $n$.
Then, the success counts $\listp{c_j}$ are binomially distributed for each $j$ with success probability $P_{n_j}$.
Here we assume that the function $P(n)$ is fixed for a given \fr{} benchmarking experiment, and for example does not depend on the experiment design.
This assumption will hold under broad experimental circumstances, even when non-Markovian, gate-dependent, or gate-position-dependent errors are present.
For example, consider a very general scenario where a different non-Markovian error process occurs at each step of a randomized benchmarking sequence.
Such a situation can occur if there is a quantum memory system $E$ and at the $k$th step there is an error channel $\Lambda_{SE}^{(k)}$ that acts on both the system $S$ that is being benchmarked and the memory $E$. 
In this scenario, if we assume that the system and environment are initialized in the same state at the beginning of each trial and that the error channels $\Lambda_{SE}^{(k)}$ are fixed, there is still a well-defined success probability at each sequence length.
The same holds true even if the error channels $\Lambda_{SE}^{(k)}$ are gate-dependent or otherwise context dependent, as long as the complete system is reset at the beginning of each trial.
If the system is reset at the beginning of each trial, then at each sequence length there is a well defined success probability, independent of the experiment design, but that may depend on the details of all the relevant error processes.
As one example of a situation where $P(n)$ can depend on the experiment design, consider a step error that depends on the total number of steps that have been performed throughout the entire history of the experiment.
In this situation, even if the order of trials is randomized, the memory effect involving the total number of steps can cause the effective $P(n)$ for this experiment to depend on the number of total trials performed at each sequence length.


We generally assume that $P(n)$ is determined by a small number of statistical parameters $\bm{\theta}=(\theta_i)$, in which case we write $P_{\boldsymbol\theta}(n)$ to make this dependence clear.
We refer to the collection of both \(\bm{\theta}\) and $P_{\boldsymbol\theta}(n)$ as a statistical model. 
The most common statistical model for randomized benchmarking is a single exponential decay with a rate that describes the step error and a proportionality constant that describes the state preparation and measurement error.
We refer to this model as the basic model and provide a concrete definition in Eq.~\ref{eq_basic_model}.
We note that the procedures for experiment design and analysis that we describe in Sections \ref{sec_opt} and \ref{sec_statistics} hold for a general model $P_{\modelplist}(n)$.

The basic model and other models that we consider in Section~\ref{sec_other_models} can be justified under certain assumptions about the behavior of the experimental system in question.
For simplicity and completeness, we provide one such set of assumptions.
\newcounter{rb_assumptions_ctr}

\refstepcounter{rb_assumptions_ctr}(\roman{rb_assumptions_ctr})\; Each \shot{} consists of state preparation of a nominal computational basis state $\ketbra{\psi_\tin}$ followed by a random sequence of steps from a two-design \cite{Dankert_two_design,ambainis_quantum_2007}, followed by a randomized final step that returns the state to a random computational basis state, followed by measurement in the computational basis. For discussion of randomized final steps and measurements see Ref.~\cite{meier_thesis}. In short, randomizing the final step and measurement allows the combined effect of state preparation and measurement errors to be treated as a single depolarizing error channel.

\refstepcounter{rb_assumptions_ctr}(\roman{rb_assumptions_ctr})\; In every \shot{} the $k$th step has an error channel $\Lambda_k$ that does not depend on the gate that the step nominally implements. 
The assumption that errors can be modeled by a channel $\Lambda_k$ is the Markovian assumption, according to the definition in Ref.~\cite{PhysRevLett.120.040405}.
The assumption of gate-independent errors is discussed in further detail in Ref.~\cite{wallman_randomized_2018}.

\refstepcounter{rb_assumptions_ctr}(\roman{rb_assumptions_ctr})\; The system is completely reset after each \shot{} and no memory effects are present between \shots{}.
This assumption disallows, for example, the possibility of step errors that depend on the temperature of a trapped ion motional mode \cite{PhysRevA.62.022311} that heats over time.
However, this assumption still allows for the possibility that step errors can be drawn from a distribution independently for each \shot{}, or can increase throughout a sequence.

When these assumptions are made, the error channels $\Lambda_k$ are `twirled' by the random gates from a two-design and become effective depolarizing channels \cite{meier_thesis}. In general, the parameters of these depolarizing channels can randomly fluctuate trial-to-trial or can depend on the gate index $k$.
The success probability of a sequence of length $n$ is determined by the composition of all the depolarizing channels at gate indices less than $n$, which can lead to more complicated behavior than a single exponential decay.
To justify the single exponential decay in the basic model, we add a final assumption.

\refstepcounter{rb_assumptions_ctr}(\roman{rb_assumptions_ctr})\; The error channels $\Lambda_k$ are independent of time and independent of the step index $k$. This assumption ensures that each effective depolarizing channel has the same depolarizing parameter.

Although these assumptions may seem restrictive, a single exponential decay can still be a good model in many situations where gate-dependent or certain time-dependent errors are present \cite{wallman_randomized_2018,PhysRevA.89.062321}.
In the case of gate-dependent errors, the observed rate of exponential decay may differ from the average fidelity of the gates relative to a fixed basis \cite{proctor.PhysRevLett.119.130502}.
The observed exponential decay rate is still indicative of gate performance, however  \cite{proctor.PhysRevLett.119.130502}.
In the case of errors that depend on a classically fluctuating quantity like temperature, randomizing the order of all \shots{} leads to a success probability at each sequence length that is averaged over the fluctuating quantity.
To good approximation, this behavior can lead to an effective model where the step errors randomly fluctuate trial-to-trial independently.
For further information about error models and assumptions in randomized benchmarking, we refer to Refs.~\cite{Figueroa_Romero_2021,proctor.PhysRevLett.119.130502,wallman_randomized_2018,PhysRevA.89.062321}.

We now provide notation and conventions for the basic model.
We use the standard definition that a depolarizing channel $\Phi$ on a Hilbert space of dimension $D$ with a depolarizing parameter $\lambda$ maps an input state $\rho$ to $\Phi(\rho) = (1-\lambda)\rho + \lambda I/D$ where $I$ is the identity operator on Hilbert space and $\lambda$ satisfies $0 \leq \lambda \leq 1 + 1/(D^2-1)$.
If a system is initialized in a pure state $\ketbra{\psi}$ and a depolarizing channel with parameter $\lambda$ is applied, the fidelity $f$ of the output state with the input state is
\begin{equation}
  f = \trb{\Phi(\ketbra{\psi})\cdot\ketbra{\psi}} = 1-\lambda + \lambda/D.
\end{equation}
The fidelity $f$ does not depend on the input state $\ketbra{\psi}$, and therefore the average fidelity of the depolarizing channel $\Phi$ is equal to $f$.
We refer to $\gerr = 1-f$ as the error of the depolarizing channel $\Phi$. 
If depolarizing channels with parameters $\listb{\lambda_i}$ are concatenated, the resulting channel is a depolarizing channel with parameter $\lambda = 1 - \prod_i (1-\lambda_i)$.
When the fidelity of the concatenated channel is expressed in terms of the individual errors it simplifies to the following
\begin{equation}
  f = \frac{1}{D} + \frac{1}{\alpha}\prod_{i}(1-\alpha \varepsilon_i),
\end{equation}
where $\alpha = \frac{D}{D-1}$.
This motivates the following definition of the basic model,
\begin{equation}
  \label{eq_basic_model}
  P_{\boldsymbol\theta}(n) = \frac{1}{D} + \frac{1}{\alpha}(1-\alpha \theta_0)(1- \alpha \theta_1)^n,
\end{equation}
where $n$ is the sequence length, $\theta_0$ is the SPAM error, and $\theta_1$ is the step error.
Here $n$ is a non-negative integer, and $\theta_0,\theta_1 \in [0,1]$.
In Section~\ref{sec_other_models} we describe several other models of experimental interest that generalize the basic model. 
An important property of the basic model is that the SPAM parameter $\theta_0$ appears affine linearly in the expression for $P_{\boldsymbol\theta}(n)$.
As a result, a randomly fluctuating SPAM parameter is indistinguishable from a constant SPAM parameter equal to the mean of the distribution of random fluctuations.
Fully-randomized benchmarking is therefore completely insensitive to drifts in the SPAM parameter, as long as the drifts are uncorrelated with the choice of sequence lengths.
In contrast, in a non-fully-randomized experiment, drifting SPAM parameters can cause an increase in the variance of the sequence fidelities over random sequences, which can lead to larger total uncertainties.

}
\section{Optimized experiment design for \fr{} benchmarking}
\label{sec_opt}

We describe a procedure to optimize the design of a \fr{} benchmarking experiment for statistical performance according to an arbitrary pre-chosen statistical model $P_{\boldsymbol\theta}(n)$.
The goal of the optimization is to minimize the anticipated uncertainty of the inference of the parameter of interest, $\prechosenparam$. In many cases the parameter of interest is the step error $\theta_1$.
The optimization is performed by linearizing the model around a reference point $\paramrefpoint$ and constructing a linear estimator for $\prechosenparam$ that has minimum variance and is insensitive to the other parameters at the reference point.
The standard deviation of the optimal linear estimator is the uncertainty of inference of $\prechosenparam$ in the linearized model, and is therefore a `first-order' approximation of the anticipated uncertainty of inference of $\prechosenparam$ in the actual model.
The accuracy of this approximation depends on the `closeness' of the reference point to the true point and on the nearby `curvature' of the statistical model.
For more details we refer to Refs.~\cite{pukelsheim_exp_design,nielsen_cr}.
The accuracy of this approximation can be checked in simulation by comparing maximum-likelihood inference in the full model to maximum-likelihood inference in the linearized model for data sampled near the reference point.
We also assume that a sufficiently accurate reference point can be obtained from prior calibration on the experimental setup.

The optimized experiment design that we describe here is called C-optimal design, and it can be formulated as a linear program \cite{c_optimal_linear_program,elfving_c_opt_design}.
Other types of optimization with different objectives are also possible.
For example, a general formulation of C-optimal design minimizes the variance of an arbitrary linear combination of model parameters.
Similarly, another objective could be to jointly minimize a weighted sum of variances of several parameters.
All of these objectives lead to convex optimization problems and have a close connection to Fisher information \cite{pukelsheim_exp_design,c_optimal_linear_program,nielsen_cr}.
For convenience, in Appendix~\ref{sec_opt_le} we provide a description of the relationship between C-optimal design and Fisher information.
For more information and details about these types of optimized experiment design, we refer to Refs.~\cite{fedorov_exp_design,pukelsheim_exp_design,sagnol_socp,c_optimal_linear_program}. 

Here we present the optimization procedure to minimize the anticipated uncertainty of a single parameter $\prechosenparam$, specifically in the context of designing experiments for \fr{} benchmarking.
The optimization is performed over the parameters $n_{j}$ and $w_{j}$
  of the experimental design, subject to a constraint on the total experimental time $T$.
Altogether, the inputs to the optimization are: the statistical model $P_{\boldsymbol\theta}(n)$, the reference point $\paramrefpoint$, the pre-chosen parameter $\prechosenparam$, the maximum sequence length $n_{\text{max}}$ that is available in an experiment, and a list $t_n$ of the amount of experiment time that it takes to experimentally perform a sequence of length $n$.
The details of the optimization procedure are as follows.
Let $P_{\paramrefpoint}(n)$ denote the success probabilities at the reference point $\boldsymbol{\theta}^{(0)}$ as a function of the sequence length $n$, and let $\delta p_n$ denote small changes in $P_{\boldsymbol\theta}(n)$ around $P_{\paramrefpoint}(n)$, so $P_{\boldsymbol\theta}(n) = P_{\boldsymbol{\theta}^{(0)}}(n) + \delta p_n$.  
Any differentiable model can be linearized around the reference point $P_{\boldsymbol{\theta}^{(0)}}(n)$. Let $L_{ni} = \frac{\partial P(n)}{\partial \theta_i}\rvert_{\paramrefpoint}$ be the gradient of the model at the reference point.
  Then we can write 
\begin{equation}
\delta p_n = \sum_i L_{ni} \delta\theta_{i},
\end{equation}
to first order in the $\delta\theta_{i}$. For the purpose of optimization we now assume
the linearized model.

Let $\delta \hat{p}_n$ denote the empirical estimator of $\delta p_n$ obtained from the observed frequency of successes after subtracting the probability of success at the reference point. If we denote the observed number of success counts by $\hat{c}_n$, we have
\begin{equation}
\delta \hat{p}_n = \frac{\hat{c}_n}{w_n} - P_{\boldsymbol{\theta}^{(0)}}(n).
\end{equation}
We consider linear estimators $\hat{A}$ of the form
\begin{equation}
  \label{eq_lin_est}
 \hat{A} = \sum_n C_n \delta \hat{p}_n,
\end{equation}
where we choose the coefficients $C_n$ so that $\hat{A}$ estimates $\delta\theta_{i_0}$ with minimum variance at the reference point.
Concretely, $\hat{A}$ estimates $\delta\theta_{i_0}$ if the coefficients satisfy $\sum_n C_n L_{ni} = \delta_{i i_0}$, which implies that $\langle \hat{A} \rangle = \delta\theta_{i_0}$ and that $\hat A$ is insensitive to the other parameters $\theta_{i\neq i_0}$.
Of the many linear estimators that satisfy these constraints, we wish to construct one with the minimum variance at the reference point, subject to the additional constraint that the experiment takes a total time $T$.
If a \shot{} with a sequence of length $n$ takes a time $t_n$, then this constraint can be expressed as $\sum_n w_n t_n = T$.
At the reference point the variance $v_{n}$ of $\delta \hat{p}_n$ is determined by the number of \shots{} $w_n$ and the binomial statistics of a single \shot{} according to
\begin{equation}
  v_n = \text{var}\,\delta\hat{p}_n =  \frac{P_{\boldsymbol{\theta}^{(0)}}(n)\pqty{1-P_{\boldsymbol{\theta}^{(0)}}(n)}}{w_n}.
\end{equation}
It follows that the variance $V$ of $\hat{A}$ satisfies
\begin{equation}
  \label{eq_lin_est_var}
V = \text{var}\,\hat{A} = \sum_n C_n^2 \frac{v_n}{w_n},
\end{equation}
where we have used the independence of $\delta \hat{p}_n$ for different $n$.
In total, to construct the optimal linear estimator we minimize $V$ jointly over the $C_n$ and the $w_n$, subject to the constraints $\sum_n C_n L_{ni} = \delta_{ii_0}$ and $\sum_n w_n t_n = T$.
We are free to optimize over the $C_n$ and the $w_n$ in either order.
The optimization of the $w_n$ at fixed $C_{n}$ yields a closed form solution, which can then be optimized over choices of the $C_{n}$ by a linear program as explained in the following paragraph.

  We now fix the \(C_{n}\) and minimize \(V\) over choices of the
  $w_{n}$ subject to the constraint $\sum_n w_n t_n =
  T$. For this we introduce the Lagrange multiplier \(\lambda\) and
  find the critical points with respect to $w_{n}$ of
  \begin{equation}
    V_{\lambda} = \sum_n C_n^2 \frac{v_n}{w_n} + \lambda \left(\sum_n w_n t_n - T\right).
  \end{equation}
  Differentiating by \(w_{n}\) and solving for \(w_{n}\) gives the critical point equations
  \begin{equation}
    \label{eq_opt_weights}
    w_n = \frac{\abs{C_n}\sqrt{v_n}}{\sqrt\lambda \sqrt{t_n}},
  \end{equation}
  which we substitute back into the expression for \(V_{\lambda}\) to obtain
  \begin{align}
    V_{\lambda,\text{opt}} &=
                             \sum_n |C_n| \sqrt{\lambda}\sqrt{v_nt_{n}} +
                             \lambda\left(\sum_n
                             \frac{|C_{n}|\sqrt{v_{n}t_{n}}}{\sqrt{\lambda}} -
                             {T}\right)\nonumber\\
                           &=2\sqrt{\lambda}\sum_n |C_n| \sqrt{v_nt_{n}} - \lambda T.
  \end{align}
  Substituting the solution for \(w_{n}\) into the constraint and rearranging terms
  constrains \(\lambda\) according to
  \({\lambda}  = \left({\sum_n \abs{C_n}\sqrt{v_n t_n}}/{T}\right)^{2}\). Substituting
  this value for \(\lambda\) into the expression for \(V_{\lambda,\text{opt}}\) gives
  the minimum variance for fixed $C_{n}$
  \begin{align}
    \label{eq_vopt}
    V_{\text{opt}} &= \frac{1}{T} \left(\sum_{n}\abs{C_{n}}\sqrt{v_n t_n}\right)^{2}.
  \end{align}
  To minimize \(V_{\text{opt}}\) over the constrained values of
  $C_{n}$, it suffices to minimize the quantity
  \(F=\sum_{n}\abs{C_{n}}\sqrt{v_n t_n}\) with the linear constraints
 $\sum_n C_n L_{ni} = \delta_{i i_0}$. This can be 
  done by means of a linear program using a standard method for handling
  the absolute values~\cite{convex_boyd}. The resulting linear program is
  \begin{alignat}{3}
    \textrm{Minimize:\ }&F=\sum_{n}\tilde C_{n}\sqrt{v_n t_n}\notag\\
    \textrm{Variables:\ }& \listp{C_{n}}_{n=1}^{n_\text{max}},\listp{\tilde C_{n}}_{n=1}^{n_\text{max}}\notag\\
    \textrm{Subject to:\ }& \textrm{for all \(n\),}\tilde C_{n}\geq 0 ,\notag\\
    &\textrm{for all \(n\)}, -\tilde C_{n}\leq C_{n}\leq \tilde C_{n},\notag\\
    &\sum_{n}C_{n} L_{ni}=\delta_{ii_0} .
  \end{alignat}
Once the optimal $C_n$ are determined, the optimal $w_n$ can be determined by substitution into Eq.~\ref{eq_opt_weights}.
After this substitution the optimal $w_n$ will be non-negative real numbers, and must be rounded to integer values to design a real experiment. 
In practice the rounding has only a small effect on the statistical power of the experiment.
In total, this optimization method determines the experiment design that has the minimum variance of the best linear estimator of the parameter $\prechosenparam$ in the linearized model at the reference point. 
This variance can be computed in terms of the optimal $C_n$ according to Eq.~\ref{eq_vopt}.
As we describe in Section~\ref{sec_statistics}, for analysis of randomized benchmarking data we use the maximum likelihood-estimator in the full model.
In the limit of a large amount of collected data we expect the variance of the maximum likelihood-estimator in the full model to match the variance of the best linear estimator of $\prechosenparam$ in the linearized model.
In any realistic scenario discrepancies can arise between the two variances as a result of the finite amount of collected data, or because the reference point used for the optimization differs from the true point.
In this sense, the anticipated variance of the optimal experiment design in Eq.~\ref{eq_vopt} should be regarded as approximate, although we expect good agreement in well-behaved cases.
For example, in the randomized benchmarking experiments run at NIST that we describe in Section~\ref{sec_exp_implementation}, we find that the anticipated variance closely matches the observed variance.

\newcommand{\fntheta}{P(n)_{\{\theta_i\}}}
\section{Models of Randomized Benchmarking}
\newcommand{\probabs}{\tilde\sigma}
\label{sec_other_models}
Here we consider several models of \fr{} benchmarking that generalize the basic model.
As we describe in Section~\ref{sec_overview}, a model of \fr{} benchmarking is specified by statistical parameters $\boldsymbol\theta$ and a function $P_{\boldsymbol{\theta}}(n)$ that determines the success probability at each sequence length $n$ in terms of those parameters.
Different models of \fr{} benchmarking can be useful to experimentalists in different scenarios.
As a concrete example, consider a situation where the step error increases throughout a sequence due to ion heating.
In this case, the heating rate can be treated as a statistical parameter contained in $\boldsymbol\theta$, and one can calculate $P_{\boldsymbol\theta}(n)$ in terms of the heating rate and the step error.
Then, the methods of Section~\ref{sec_opt} can be used to design an experiment to maximize information about the heating rate according to this model.
However, in many cases an experimentalist may have incomplete information about the time-dependent behavior, or in fact may believe that there is no noticeable time-dependent behavior. 
In such a situation it is beneficial to consider a very general model that can capture a wide range of time-dependent behavior.
Such a model can be used to design experiments to detect potential deviations from exponential decay, or to provide evidence that no noticeable time dependence or non-Markovianity is present.
As one candidate for such a model, we introduce the `moments model' in the following paragraph.

To motivate the moments model, we consider a step error that is constant
  throughout the random sequence of an individual \shot{}, but is drawn independently from a probability distribution
  $\probabs(\varepsilon)$ for each \shot{}.
  This behavior can occur if there are classically-fluctuating quantities in an experiment that affect the step error on time-scales slow compared to the duration of a single sequence.
  One such example is temperature fluctuations in the apparatus that performs the physical pulses that make up the gates.
  Furthermore, when the order of all \shot{}s is randomized, even a monotonically-increasing error rate throughout the entire experiment will behave as though the error is drawn from a distribution, as long as the rate of the error increase is  slow compared to the duration of a single sequence.
  Accordingly, the
  success probability \(P(n)\) in such a situation is
  \begin{equation}
    P(n) = \frac{1}{D} + \frac{1}{\alpha}(1-\alpha \theta_0)\int d\varepsilon \probabs(\varepsilon) (1- \alpha \varepsilon)^n,
  \end{equation}
  where $\alpha = \frac{D}{D-1}$ and the parameter $\theta_0$ describes the SPAM error. 
  As written, this model is parametrized by \(\theta_{0}\) and \(\probabs\), which is an infinite dimensional
  parameter. 
  Below we show that only \(N+1\) parameters are relevant if
  the sequence length is bounded by \(N\).
  The basic model corresponds to the case of \(\probabs(\epsilon)=\delta(\epsilon-\theta_{1})\), where \(\theta_{1}\) is the step error and $\delta$ denotes the Dirac delta distribution.
  For a general distribution $\probabs(\varepsilon)$, we denote the mean of $\probabs(\varepsilon)$ by $\theta_1$ and interpret it as the parameter analogous to step error.
At times it is convenient
  to shift the probability distribution \(\probabs(\varepsilon)\) by its mean $\theta_1$. We define \(\sigma(\varepsilon)=\probabs(\varepsilon+\theta_{1})\) so that
  \begin{equation}
    P(n) = \frac{1}{D} + \frac{1}{\alpha}(1-\alpha \theta_0)\int d\varepsilon \sigma(\varepsilon)(1- \alpha \theta_1 - \alpha \varepsilon)^n.
  \end{equation}
  The parameters of the model are now \(\theta_{0}, \theta_{1}\), and \(\sigma\),where the probability distribution \(\sigma\) is constrained to have mean \(0\) and support in \([-\theta_{1},1-\theta_{1}]\). 
  The basic model is recovered with \(\sigma(\varepsilon) = \delta(\varepsilon)\). 
Applying the binomial expansion to the \(n\)'th power in the expression for \(P(n)\) gives
\begin{equation}
P(n) = \frac{1}{D} + \frac{1}{\alpha}(1-\alpha \theta_0)\pqty{(1-\alpha\theta_1)^n + \sum_{k=2}^n \binom{n}{k}(1-\alpha\theta_1)^{n-k}(-\alpha)^k\int d\varepsilon \sigma(\varepsilon) \varepsilon^k},
\end{equation}
where the $k=1$ term vanishes by the assumption that $\sigma(\varepsilon)$ has mean $0$.
This motivates the introduction of the moment parameters $\theta_k := \int d\varepsilon \sigma(\varepsilon) \varepsilon^k$ for $k = 2$ to $N$.
In terms of these parameters the success probability can be written
\begin{equation}
P_{\boldsymbol\theta}(n) = \frac{1}{D} + \frac{1}{\alpha}(1-\alpha \theta_0)\pqty{(1-\alpha\theta_1)^n + \sum_{k=2}^n \binom{n}{k}(1-\alpha\theta_1)^{n-k}(-\alpha)^k\theta_k}.
\end{equation}
We refer to this model as the `moments model' and we refer to the parameters $\theta_k$ for $k \geq 2$ as the moments parameters. 
For practical use, the moments parameters are truncated for $k$ larger than some $k_\text{max}$, so that $\theta_k = 0$ for $k > k_\text{max}$.
For example, when we make certain comparisons to published experiments in Section~\ref{sec_comparisons}, we use the moments model with two non-zero moments parameters $\theta_2,\theta_3$ for a total of four parameters.
When we design the experiments in Section~\ref{sec_exp_implementation} we also use the moments model with four total parameters.
When we analyze those experiments we use the moments model with three total parameters, where we remove $\theta_3$.
In that case, we report $\sqrt{\theta_2}$ because this is on the same scale as $\theta_1$ and is the standard deviation of $\sigma$ if $\theta_2$ comes from a true probability distribution.
We note that if all the moments parameters are zero, the moments model reduces to the basic model with spam error $\theta_0$ and step error $\theta_1$.
We note that the parameters
  \(\theta_{2},\ldots\) are the mean-subtracted moments of the original distribution \(\probabs\). 
  For later use, we denote the moments of \(\probabs\) as \(\tilde\theta_{k}=\int d\varepsilon \varepsilon^{k}\probabs(\varepsilon)\). 
  Treating \(\theta_{1}\) as a 
  constant, for \(k\geq 2\), \(\theta_{k}\) is an affine linear combination of \(\tilde\theta_{2},\ldots \tilde\theta_{k}\), and similarly for \(\tilde\theta_{k}\) in terms of
  the \(\theta_{2},\ldots,\theta_{k}\).

The moments model is universal in the following sense.
In the absence of \shot{}-dependent step errors, the most general benchmarking
  model has arbitrary success probabilities \(P(n)\) depending on \(n\).
  We show that any such success probabilities can be modeled by a suitable choice
  of parameters of the moments model, provided the implicit linear restrictions
  on the moment parameters due to positivity and support constraints of the
  probability distribution \(\sigma\) are lifted.
We first write the moments model in terms of
the moments of \(\probabs\),
\begin{equation}
  \label{eq_moments_model_absolute}
  P(n) = \frac{1}{D} + \frac{1}{\alpha}(1-\alpha \theta_0)\pqty{1 + \sum_{k=1}^n \binom{n}{k}(-\alpha)^k\tilde\theta_k}.
\end{equation}

The parameter \(\theta_{0}\) linearly determines and is determined
by \(P(0)\).  For the remaining probabilities, we fix
\(\theta_{0}\).  Eq.~\ref{eq_moments_model_absolute} establishes a linear relationship
between the \(P'(n) =P(n)-P(0)\) for \(n\geq 1\) and the
\(\tilde\theta_{k}\) for \(k\geq 1\) of the form
\(P'(n) = \sum_{k\geq 1}M_{nk}\tilde \theta_{k}\). The matrix
\(M_{nk}\) is lower triangular with diagonal entries
\(M_{nn}=(P(0)-1/D)(-\alpha)^{n}\). 
Here we assume that $P(0) \neq 1/D$. If $P(0)=1/D$, then the initial state would be completely depolarized and the choice of moment parameters would be irrelevant. 
With this assumption the diagonal entries $M_{nn}$ are nonzero, and $M$ therefore has a lower triangular inverse.
It follows that in the absence of constraints on
the \(\tilde\theta_{k}\), all possible \(P(n)\) can be modeled with
a choice of the moment parameters. If the maximum sequence length
under consideration is $n_\text{max}$, we can truncate the matrix at \(n=n_\text{max}\)
and model \(P(0),\ldots P(N)\) with a choice of
\(\tilde\theta_{1},\ldots \tilde \theta_{N}\), or equivalently
\(\theta_{1},\ldots,\theta_{N}\), for any fixed \(\theta_{0}\).

In addition to the
basic model and the moments model, another model of experimental
interest is one where the errors in a sequence experience drift as a function of position within the sequence.
To motivate this behavior we consider a miscalibrated single-qubit gate where the miscalibration drifts linearly as a function of time but is reset at the beginning of each sequence.
Concretely, we consider a gate $U$ that nominally implements a $\pi$ rotation about the $x$-axis of the Bloch sphere and can be expressed as $U = \bexp{-i (\pi/2) X}$, where $X$ is the Pauli-$X$ operator.
We denote the action of the possibly miscalibrated gate by
  $\tilde{U} = \bexp{-i (\pi/2 + \phi)X}$,
where $\phi$ is an error parameter that describes the angle of erroneous rotation.
The average fidelity of $\tilde{U}$ with the nominal gate $U$ is equal to $1/3 + 2\cos^2(\phi)/3$.
A plausible error model is that the erroneous rotation $\phi$ depends linearly on time, which could correspond physically to a linear drift of Rabi frequency.
If the gate is perfectly calibrated at $t=0$, expanding to lowest order for short times shows that the error will grow quadratically.
If $\phi\neq0$ at $t=0$ and the expansion is performed to second order, the error will in general have both linear and quadratic dependence for short times.
Altogether, this motivates consideration of the following approximate model,
\begin{equation}
  \label{eq_lin_quad}
  P(n) = \frac{1}{D} + \frac{1}{\alpha}(1-\alpha \theta_0)\pqty{\prod_{k=1}^n (1-\alpha(A + Bk + Ck^2))},
\end{equation}
where $\theta_0$ is a SPAM parameter
and $A,B,C$ are parameters that govern the linear and quadratic drift.
One question of possible experimental relevance is whether this drift model can be distinguished from the moments model when the error distribution is restricted to be a true probability distribution $\probabs(\varepsilon)$. 
Here we show that this is indeed possible for at least one region of the space of parameters.
In particular, we consider $\theta_0 = 0, C=0$, and approximate Eq.~\ref{eq_lin_quad} to lowest order in $B$ for $B > 0$. 
When we determine the moments parameters that match this model we find that $\theta_2 < 0$ for this region of parameter space, which is impossible for a true second moment.
To determine the matching moments parameters, we follow the procedure outlined in the previous paragraph.
In the approximate linear drift model we have $P(1) = 1/D + (1 - \alpha A - \alpha B)/\alpha$ and in the moments model we have $P(1) = 1/D + (1-\alpha\theta_1)/\alpha$. 
We equate these to determine $\theta_1$ and find $\theta_1 = A + B$.
Similarly, we then equate $P(2)$ in both models
\begin{equation}
  (1-\alpha\theta_1)(1-\alpha\theta_1-\alpha B) = (1-\alpha\theta_1)^2 + \alpha^2\theta_2.
\end{equation}
Solving for $\theta_2$ we find $\theta_2 = -B(1-\alpha\theta_1)/\alpha$, which satisfies $\theta_2 < 0$ when $B > 0$.
Equating $P(3)$ in both models we find that $\theta_3$ is of order $O(B^2)$, and by induction we find that $P(k)$ is of order $O(B^2)$ or higher when $k \geq 3$.
We conclude that the parameters $\theta_k$ for $k \geq 3$ can be dropped to good approximation when $B$ is small.
In total, we conclude that if the true model is the linear drift model for small positive $B$, an analysis using the moments model with parameters $\theta_0,\theta_1,\theta_2$ would likely find $\theta_2 < 0$ in the large data limit, which is impossible for a true second moment. 

Altogether, the results of this section motivate the potential strategy for experimentalists interested in running randomized benchmarking in the presence of unknown sources of time dependence or non-Markovianity.
First, use the moments model with four total parameters to design a \fr{} benchmarking experiment, by using the anticipated step error and SPAM errors to construct the reference point and setting the reference values of the higher moments to zero.
Then, analyze the experiment according to both the basic model and the moments model with three total parameters to look for evidence of time dependence or non-Markovianity.
If the basic model is a good fit to the observed data, then one can conclude that there is no significant evidence of time dependence or non-Markovianity. However, this is not necessarily evidence that no time dependence or non-Markovianity is present.
If the basic model is a poor fit, but the moments model with three total parameters is a good fit, then the sign of the second moment can provide some hint as to the source of possible time dependence.
Concretely, a positive second moment hints at `slow' drift compared to the durations of experimental trials, while a negative second moment hints at `fast' drift compared to the durations of experimental trials.
In this case, as well as the case where the moments model with three total parameters is still a poor fit, it is prudent for the experimentalist to investigate the physical origin of the time dependence or non-Markovianity that is causing the model deviation. This paper's formalism can be adapted to other strategies depending on the types of errors present in the system being studied.

\section{Analysis of achievable uncertainty improvements}
\label{sec_comparisons}
\newcounter{tcol}

We illustrate the advantages of full randomization by comparing the uncertainties achieved in several published experiments to the uncertainties that could have been achieved with \fr{} benchmarking with the same experiment design, meaning the same sequence lengths and same number of total trials at each sequence length.
We also demonstrate that additional improvements in uncertainty could have been achieved by optimizing the experiment design according to the procedure in Section~\ref{sec_opt}.
The published randomized-benchmarking experiments that we use for specific comparisons are Refs.~\cite{brown_single-qubit-gate_2011,PhysRevLett.125.080504,huang_fidelity_2019}.
For each past experiment we assume that the basic model in Eq.~\ref{eq_basic_model} is accurate and we use the procedure in Appendix~\ref{sec_opt_le} to construct the optimal linear estimator for step error according to the reported sequence lengths and reported total number of \shots{} at each sequence length, and using the reported step error and SPAM error as the reference point.
In all of these experiments the same random sequences were repeated many times, but our construction of the optimal linear estimator assumes that the experiment was \fr{} and that a new random sequence was drawn for each \shot{}.
Therefore, we interpret the standard deviation of the optimal linear estimator as the anticipated uncertainty if the experiment had been \fr{}, and we compare it to the uncertainty actually reported by each experiment.
Then, we run the optimization described in Section~\ref{sec_opt} to construct the optimal experiment design according to the basic model.
For Refs.~\cite{huang_fidelity_2019,PhysRevLett.125.080504} we assume that the step time is equal to the SPAM time and for Ref~\cite{brown_single-qubit-gate_2011} we assume that the step time is $100$ times smaller than the SPAM time.
We interpret the anticipated uncertainty returned by the optimization as the size of the confidence interval for each experiment if it had been \fr{} and the optimal experiment design had been used.
Finally, we repeat the optimization for the four-parameter moments model to see how the anticipated uncertainty is affected by a more general model.
All of these observations are recorded in Table.~\ref{fig_comparisons}.
We generally observe that improvements in uncertainty are possible both from fully randomizing the experiment and from using the optimal experiment design.

\begin{table}[h]
  \caption{Results of a numerical signal-to-noise comparison between past randomized benchmarking experiments and experiments optimized according to the procedure in Section~\ref{sec_opt}.
  The columns show the referenced benchmarking experiment; the gate error and uncertainty reported by each experiment;
  the anticipated uncertainty for a \fr{} experiment with the reported experiment design; obtained as in Appendix~\ref{sec_opt_le};
  and the expected uncertainty if the experiment design is optimized, as described in Section~\ref{sec_opt}, for the basic model and the four-parameter moments model respectively.}
  \label{fig_comparisons}
\begin{center}
  \begin{tblr}{colspec={|Q[c]|Q[c]|Q[c]|Q[c]|Q[c]|Q[c]|},rowspec={|Q[m]|Q[m]|Q[m]|Q[m]|},colsep=1mm,rowsep=1mm}
 {Experiment\refstepcounter{tcol}\label{tcol_experiment}} & {Reported \\ step error\refstepcounter{tcol}\label{tcol_reported_err} }& {Reported \\ uncertainty \refstepcounter{tcol}\label{tcol_reported_std}} &
 {Fully-randomized \\ anticipated \\ uncertainty \\ (basic model)\refstepcounter{tcol}\label{tcol_fr} }&
 {Optimized \\ anticipated \\ uncertainty \\ (basic model)\refstepcounter{tcol}\label{tcol_opt_basic} }&
 {Optimized \\ anticipated \\ uncertainty \\ (moments model)\refstepcounter{tcol}\label{tcol_opt_moments}}\\
 Ref.~\cite{brown_single-qubit-gate_2011} & $2.0 \times 10^{-5}$ & $2 \times 10^{-6}$ & $2.1 \times 10^{-6}$ & $1.0 \times 10^{-6}$ & $1.6 \times 10^{-6}$\\
 Ref.~\cite{PhysRevLett.125.080504} & $8.3 \times 10^{-3}$ & $2 \times 10^{-4}$ & $1.2 \times 10^{-4}$ & $1.1 \times 10^{-4}$ & $1.7 \times 10^{-4}$\\
 Ref.~\cite{huang_fidelity_2019} & $5.3 \times 10^{-2}$ & $4 \times 10^{-3}$ & $8.8 \times 10^{-4}$ & $4.3 \times 10^{-4}$ & $1.3 \times 10^{-3}$\\
\end{tblr}
\end{center}
\end{table}
In the case of Ref.~\cite{huang_fidelity_2019}, we observe more than a factor of four improvement in anticipated uncertainty if the experiment is \fr{}.
As we show in Section~\ref{sec_fr_variance}, the size of the improvement in uncertainty from fully randomizing depends on the true error model, and is larger if the true errors are closer to unitary errors.
For the parameters reported in Ref.~\cite{huang_fidelity_2019} we find that an improvement of this size from fully randomizing is possible if the true errors are unitary.
Further details of this comparison are in Section~\ref{sec_fr_variance}.

\begin{figure}
  \captionsetup[subfloat]{labelfont=bf, farskip=0mm}
  \subfloat[]{
  \includegraphics[width=.5\textwidth]{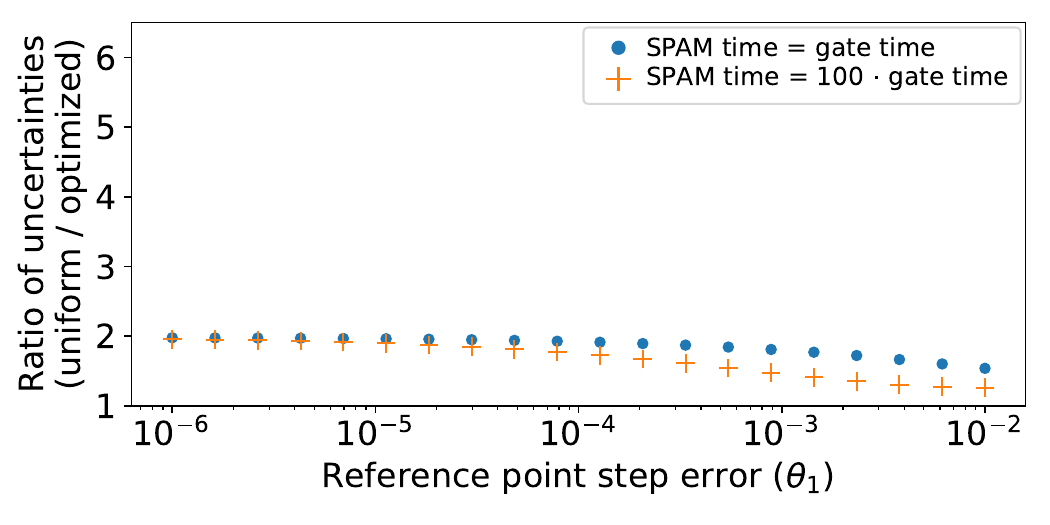}%
  }%
  \subfloat[]{
  \includegraphics[width=.5\textwidth]{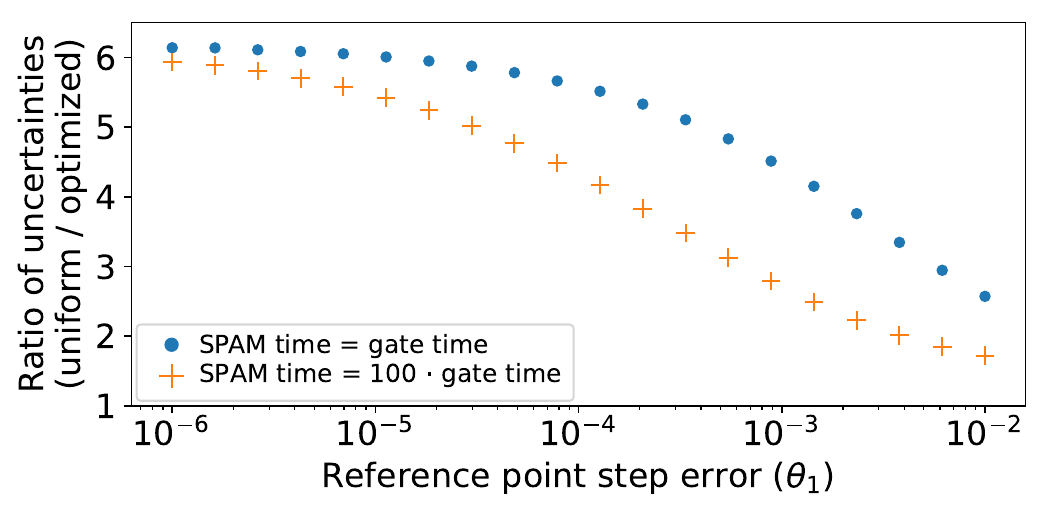}%
  }
  \caption{Comparison between hypothetical \fr{} experiments that use either an optimized experiment design or a uniform design of evenly-weighted sequence lengths.
    The plots show the ratio of the anticipated standard deviations of the step error (uniform/optimized), as a function of the reference step error. In all cases, the SPAM parameter $\theta_0$ is fixed at $10^{-2}$ and the total experiment time set to a constant.
    For the uniform experiment design, $20$ evenly-spaced sequence lengths from $[1,1/\theta_1]$ are used and the number of trials at each length is the same. In both plots, the dots show the comparison assuming that the SPAM time is equal to the step time and the plus signs show the comparison assuming that the SPAM time is larger than the step time by a factor of $100$.
    Larger ratios indicate a larger benefit from optimizing the experiment design.
    In plot (a) we use the basic model and in plot (b) we use the moments model with four total parameters.
    }
  \label{fig_uniform_comparison}
\end{figure}

In addition, we numerically explore the improvement obtained by optimizing the experiment design for hypothetical \fr{} experiments. 
The comparisons are made between uniform experiment designs where the sequence lengths are chosen uniformly in a fixed range and the same number of trials are performed at each sequence length, and a optimized experiment designs constructed according to the method in Section~\ref{sec_opt}.
The optimized experiment designs are constrained to take the same total time as the corresponding uniform experiments.
To compare uniform experiment designs to optimized designs, we compute the standard deviations of the optimal linear estimator in the linearized model, as described in Section~\ref{sec_opt} and Appendix~\ref{sec_opt_le}, and take the ratio of these standard deviations.
Larger ratios indicate a larger benefit from optimizing and the square of this ratio corresponds to the ratio of experiment times required to achieve the same standard deviation.
The results of these comparisons are shown in Fig.~\ref{fig_uniform_comparison}. In plot (a) we use
 the basic model with the spam error parameter $\theta_0$ set to $10^{-2}$ and step error $\theta_1 \in \left[10^{-6},10^{-2}\right]$. 
 In plot (b) we use the moments model with four total parameters, where the reference values of the moments parameters are set to zero, $\theta_0$ is set to $10^{-2}$, and $\theta_1$ ranges over $\left[10^{-6},10^{-2}\right]$.
 In both plots the uniform experiment design consists of $20$ uniformly spaced sequence lengths in the range $[1,1/\theta_1]$. 
 The ratio of the SPAM time to the step time is either set to $1$ or $100$ and both options are shown in the plots.
At a step error of $10^{-6}$ and when the ratio of the SPAM time to the step time is $100$, we observe a reduction in standard deviation by a factor of $1.96$ for the basic model and by a factor of $5.9$ for the moments model with four parameters.
These improvements correspond to time savings by factors of $3.8$ and $35.2$ respectively.
\section{Variance analysis of \fr{} benchmarking}
\label{sec_fr_variance}

  For randomized benchmarking with a fixed set of sequence lengths and a fixed number of \shots{} at each sequence length, \fr{} benchmarking generally yields lower uncertainty
  than randomized benchmarking with multiple trials for each chosen sequence. 
  The
  uncertainty reduction depends on the error channels and is due to
  the sequence-dependent success probabilities for error channels that
  are not depolarizing. The reduction is particularly pronounced for
  unitary error channels and may be analyzed by fixing
  the sequence length $n$ and evaluating the variance of the empirical
  estimate of $P(n)$ for the general scenario where we run $M=kl$
  independent trials consisting of $k$ randomly chosen sequences where
  each sequence is run $l$ times.  
  For related work on the relationship between the number of random sequences and the variance of a randomized benchmarking experiment, we refer to Refs.~\cite{Wallman_2014, granade_accelerated_2015}.
  The \fr{} scenario has
  $k=M$ and $l=1$. We assume a sequence-dependent probability of
  success $s$.  Since the sequence is chosen randomly, the probability of
  success can be considered as a random variable with probability
  measure on $s \in [0,1]$ given by $\mu(s)$ that depends on $n$, the
  two-design used, and the actual gate-error model. 
  Regardless of the actual gate-error model, there will be some $\mu(s)$ that describes the distribution of success probabilities over random sequences.
  The goal is to estimate the average probability of
  success, which is given by
  $\bar s = \langle s\rangle_{\mu}= \int d\mu(s) s$.  For
  $i=1,\ldots, k$, let $\hat c_{i}$ be the number of observed
  successes for the $i$'th sequence.  The minimum variance estimator
  for $\bar s$ is the empirical average
  $\hat s= \frac{1}{k}\sum_{i}\frac{\hat c_{i}}{l}$.  Because the
  sequences are independent and identically distributed, each $\hat c_{i}$ is identically distributed according to a random variable $C$ which is the sum of $l$ Bernoulli random variables with success probability $S$. 
  The variance of $C$ given $S$ is $l
  S(1-S)$ and the mean of $C$ given $S$ is $lS$.   The variance of $C$ can be computed
  according to the law of total variance \cite{weiss2005course} as
  \begin{align}
    \text{var}(C) &= E(\text{var}(C|S)) + \text{var}(E(C|S))\nonumber&&\\
                   &= l\langle S(1-S)\rangle_{\mu} + l^{2}\langle (S-\bar s)^{2}\rangle &=\;\;\;& l\bar s(1-\bar s) + l(l-1) \text{var}(S).\;\;\;
  \end{align}
  This expression appears, for example, in Appendix A of Ref.~\cite{granade_accelerated_2015}.
  The variance of $\hat s$ is $\frac{1}{k}\text{var}(C)/l^{2}$. 
Accordingly,
\begin{equation}
  \label{eq_var_s_general}
  \text{var}\hat{s} = \frac{\bar{s}(1-\bar{s})}{M} + \frac{(l-1)}{M} \pqty{\pqty{\int d\mu(s)s^2} - \bar{s}^2}.
\end{equation}
The second term in Eq.~\ref{eq_var_s_general} vanishes if $l=1$, so we
can interpret it as the excess variance due to not fully randomizing
and we denote it by $Z$. 
An important point is that $Z$ depends on the
exact error model via $\int d\mu(s) s^2$.  
For example, if all errors
are gate-independent depolarizing channels then the success probability
is independent of the random sequence and
$\int d\mu(s)s^2 = \bar{s}^2$, which implies that $Z = 0$.  
In contrast, if the error channel is a fixed
  unitary, $S$ depends on the particular sequence and $Z$ may be
  significant. 
  In this regard the
observed variance in success probabilities over random sequences
can provide a rough measure of the amount of coherent error in a
randomized benchmarking experiment. 
This has been observed qualitatively in Ref~\cite{Knill_2008}, where the large variance in fidelities at each sequence length is attributed to coherent errors.
For related work to distinguish coherent and incoherent errors in a randomized benchmarking experiment by inferring a quantity called the unitarity, we refer to Refs.~\cite{Wallman_2015,PhysRevA.99.012315}.
Furthermore, Ref.~\cite{PhysRevA.100.032304} studies bounds on the excess variance in terms of the fidelity and the unitary for gate-independent noise, assuming that each individual sequence is measured infinitely many times so that its success probability is known exactly. 
Ref.~\cite{PhysRevA.100.032304} also considers a modified randomized benchmarking protocol that can help mitigate some of the excess variance from not fully randomizing, which the authors indicate can be significant for the unmodified randomized benchmarking protocol. 

To better understand the possible sizes of the excess variance $Z$, we consider a specific error model involving unitary error channels.
Consider an error model where the final state $\psi$ is assumed to be equal to the target state $\chi$ with probability $\lambda$ and is a random pure state with probability $1-\lambda$. 
This error model is intended to heuristically capture the effects of unitary errors, where the final state is always a pure state.
To express $\lambda$ in terms of $\bar{s}$, we note that
\begin{equation}
  \bar{s}  = \lambda + (1-\lambda)\int d\psi_H f_\psi = \lambda + \frac{1}{D}(1-\lambda),
\end{equation}
where $d\psi_H$ denotes the Haar measure over pure states and $f_\psi$ denotes the success probability for each random pure state $\psi$.
This relationship can be inverted to solve for $\lambda$ as a function of $\bar{s}$
\begin{equation}
  \label{eq_var_lambda}
  \lambda = \frac{\bar{s} - 1/D}{1-1/D}.
\end{equation}
In order to determine the variance of $\hat{s}$ by substituting into Eq.~\ref{eq_var_s_general} for this error model we first evaluate
\begin{equation}
  \int d\mu(s) s^2  = \lambda + (1-\lambda)\frac{2}{D(D+1)},
\end{equation} 
where we have used the fact that $\int d\psi_H f_\psi^2 = \frac{2}{D(D+1)}$ (Appendix~\ref{app_excess_variance}).
Substitution for $\lambda$ according to Eq.~\ref{eq_var_lambda} leads to the final expression
\begin{equation}
  \label{eq_var_s_final}
  \text{var}\hat{s} = \frac{\bar{s}(1-\bar{s})}{M} + \frac{l-1}{M}\pqty{\frac{\bar{s} - 1/D}{1-1/D} + \frac{1-\bar{s}}{1-1/D}\frac{2}{D(D+1)} - \bar{s}^2}.
\end{equation}
  For any given experiment, it is possible to estimate the excess
  variance due to repetition of sequences by considering the
  statistics obtained at a particular sequence length. For example,
  consider a sequence length of $20$ in the experiment reported in
  Ref.~\cite{huang_fidelity_2019}. In this experiment, $D=4$, $k=51$
  and $l=125$. At a sequence length of $20$, the reported success
  probability is $0.31$ and the $95\;\%$ confidence interval
  has a total size of approximately $0.06$ as determined from
  Fig.~4a of Ref.~\cite{huang_fidelity_2019}. If the experiment had been \fr{}, we would expect a
  total size of this confidence interval of $0.023$ when analyzed according to the basic model. For comparison,
  with the unitary error model of the previous paragraph and the
  parameters reported in Ref.~\cite{huang_fidelity_2019}, the $95\;\%$ confidence interval would have had a total size of
  $0.16$.
  It is therefore possible that the increased size of the reported confidence interval relative to the anticipated confidence interval from \fr{} benchmarking can be explained by coherent errors in the actual experiment.
\section{Statistical Analysis}
\label{sec_statistics}
For inference of the model parameters $\boldsymbol\theta$ for any model $P_{\boldsymbol\theta}(n)$ we use maximum likelihood whenever it is tractable and well-behaved, which it is for all the examples we consider in this paper.
Maximum likelihood has the advantage that it is asymptotically unbiased, meaning that as the amount of collected data grows to infinity,
the inferred model parameters match the true point in parameter space.
To obtain confidence intervals on one or more parameters one may use
statistical bootstrapping, which is a method of resampling the observed data to learn how much the inferred quantities vary as the data varies.
For more information about maximum likelihood and statistical bootstrapping we refer to Refs.~\cite{efron_bootstrap,boos_introduction_2003}.
Here we provide the log-likelihood function for an arbitrary model and discuss the possibilities for obtaining confidence intervals through statistical bootstrapping.
We also discuss a statistical analysis to possibly reject the inner model(s) of a set of nested models using an empirical likelihood ratio test with statistical bootstrapping.
This is a common technique in statistics that has also been used in quantum characterization theory \cite{Nielsen_2021}.

The log-likelihood function for an arbitrary model is as follows.
The probability $L_j$ of observing $c_j$ successes out of $w_j$ \shots{} at the sequence length $n_j$ is
\begin{equation}
  L_j = \binom{w_j}{c_j} (P_{\boldsymbol\theta}(n_j))^{c_j}(1-P_{\boldsymbol\theta}(n_j))^{w_j-c_j}.
\end{equation}
The total probability is obtained by taking a
product over all sequence lengths in the list $\listp{n_j}$. It
follows that the full log-probability $\Theta$ is 
\begin{equation}
  \Theta = \sum_{j=1}^{j_\text{max}} \pqty{\log{\binom{w_j}{c_j}} + c_j \log{P_{\boldsymbol\theta}(n_j)} + (w_j-c_j)\log{(1-P_{\boldsymbol\theta}(n_j))}}.
\end{equation}
We note that the dependence on the model parameters $\boldsymbol\theta$ is entirely through $P_{\boldsymbol\theta}(n)$.

Parametric or non-parametric bias-corrected bootstrapping \cite{efron_bootstrap,efron_jackknife} can be used to obtain confidence intervals for one or more parameters.
In typical uses of bootstrapping in quantum
  characterization, the bootstrap assumptions are not satisfied, often because
  the parameters are statistically close to the boundary. As a result, the coverage
  probabilities do not closely match the nominal confidence levels used.
  Nevertheless, at moderate confidence levels, the intervals  obtained are 
  useful for interpretation but should be treated as approximate.
For more information about potential issues with bootstrap coverage probabilities we refer to \cite{qualms_bootstrap,qualms_bootstrap_revisited} and for examples and discussion in the context of quantum information science we refer to \cite{blumekohout2012robust,Bezerra2022}.
When we optimize the design of an experiment according to the procedure
in Section~\ref{sec_opt}, the anticipated uncertainty that we minimize is intended to approximate the size of confidence intervals obtained according to the Gaussian assumption, absent any boundary issues. 
However, for experiments of finite duration the confidence intervals in general do not exactly match the anticipated uncertainty, even if the reference point used for the optimization is equal to the true point.
In the limit that the experiment duration and the amount of data become large and a Gaussian model is a good approximation, the confidence interval sizes should match the anticipated uncertainty.
There can still be deviations in the large data limit if the reference point used for the optimization does not match the true point.

In some experiments there may be two or more relevant statistical models that are nested, meaning that the inner model can be obtained from the outer model by fixing some of its parameters at constant values.
In such a situation, it may be useful to perform a statistical analysis to attempt to reject the inner model.
One such method is to use an empirical likelihood ratio test with statistical bootstrapping \cite{boos_introduction_2003,PhysRevLett.117.060505,rudinger_probing_2019}.
A standard likelihood ratio test with a chi-squared analysis would be sufficient if a Gaussian model were accurate.
However, in many relevant cases the Gaussian model does not hold and this can lead to noticeable statistical issues \cite{scholten_behavior_2018}.
For this reason, one may use an empirical likelihood ratio test which we now describe.
We denote the outer model by $P_{\boldsymbol\theta,\boldsymbol\phi}(n)$, where now there are two sets of statistical parameters $\boldsymbol{\theta},\boldsymbol{\phi}$. 
The inner model is obtained by setting $\boldsymbol{\phi}$ to some particular value. 
For a given set of data a maximum likelihood analysis can be run for both models, and the ratio of  the maximum likelihood values can be computed.
Assuming the inner model is true, the distribution of likelihood ratios can be estimated, empirically, by bootstrap resampling the data according to the inner model and computing the likelihood ratio for each resampled dataset.
With this analysis one can reject the inner model at a particular confidence level, which is based on the percentile of the observed likelihood ratio within this empirical distribution of bootstrapped likelihood ratios.
This technique is similar to that proposed in Ref.~\cite{Nielsen_2021}, which also advocates for the use of statistical tests to possibly reject inner models in a sequence of nested statistical models.
The main difference is that we use statistical bootstrapping to empirically estimate the distribution of likelihood ratios instead of using a chi-squared analysis.
In our work the nested statistical models that we consider are those described in Section~\ref{sec_other_models}.
Other nested statistical models may be useful in other situations.
One other example of nested statistical models is the family of wildcard models proposed in Ref.~\cite{blumekohout2020wildcarderrorquantifyingunmodeled}, which allows the success probabilities to deviate from exponential decay by different amounts quantified by total variation distance.
Fits to the wildcard models can be used as a diagnostic of how much the original model must be weakened in order to fit observed data.
Regardless of the particular nested models used, one key benefit of fully randomizing is that maximum likelihood inference can be used, which makes statistical tests based on likelihood ratios possible.

\begin{figure}
  \captionsetup[subfloat]{labelfont=bf, farskip=0mm}
  \subfloat[Basic model vs Moments model] {%
  \includegraphics[width=.5\textwidth]{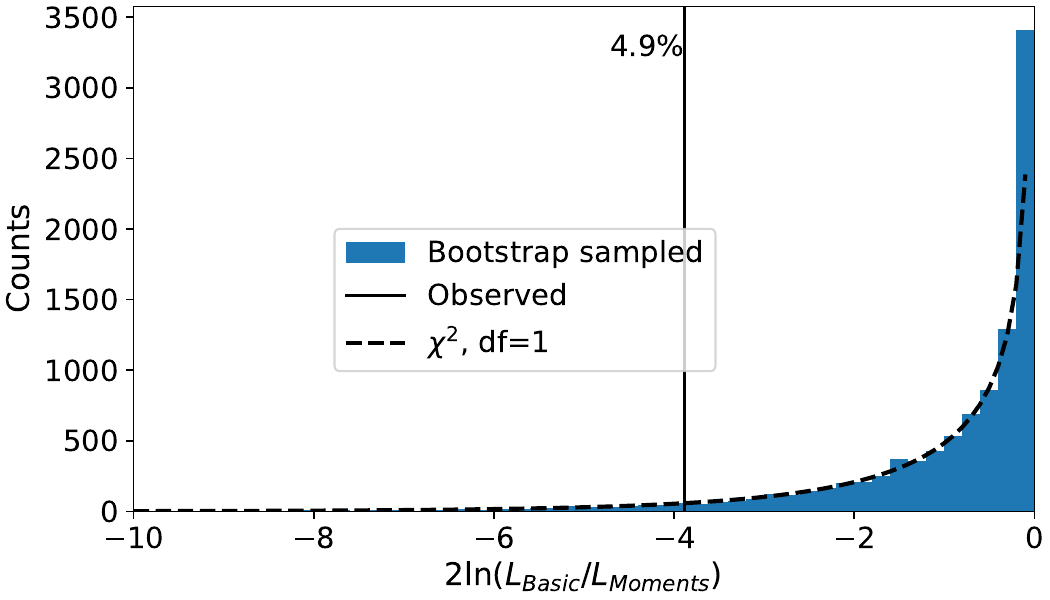}%
  }
  \subfloat[Moments model vs General model] {%
    \includegraphics[width=.5\textwidth]{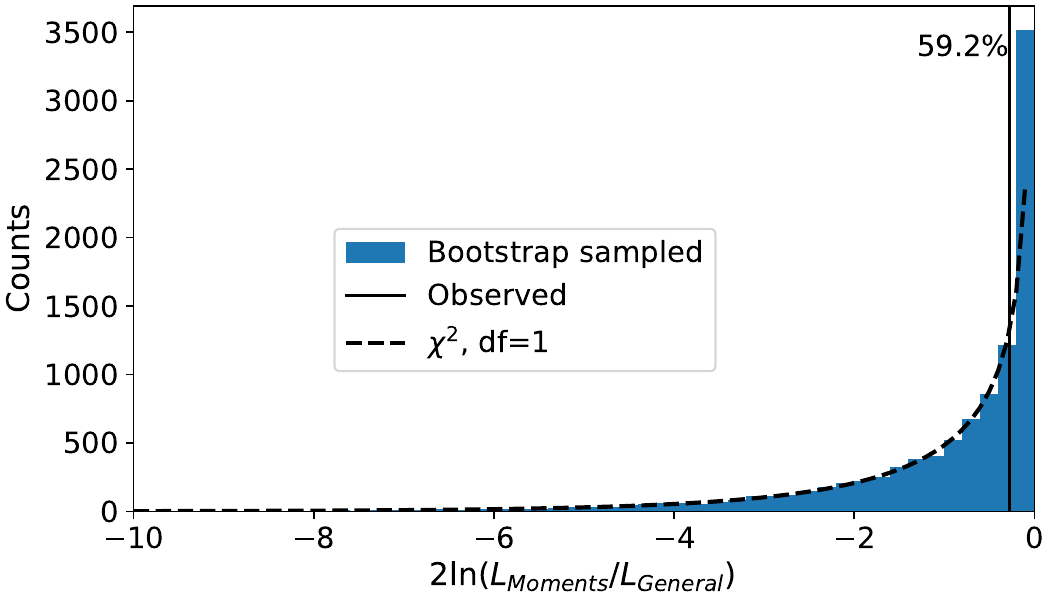}%
  }
  \caption{Results of a simulated empirical likelihood ratio test of three nested statistical models of \fr{} benchmarking.
  The three statistical models that we consider are the basic model, the moments model, and the general model. Further details about these models are in Section~\ref{sec_other_models}.
  The test is conducted
   using a simulated dataset obtained by drawing the step error $\theta_1$ independently for each \shot{} from a Gaussian distribution with mean $1\times10^{-4}$ and standard deviation $2.5\times 10^{-5}$. 
  To perform the test we use the procedure described in Section~\ref{sec_statistics} in two cases, the first where the inner model is the basic model and the outer model is the three-parameter moments model, and the second where the inner model is the three-parameter moments model and the outer model is the general model.
  In the first case the p-value to reject the basic model $4.9\%$, and in the second case the p-value to reject the three-parameter moments model is $59.2\%$.
  Thus, we find some evidence to reject the basic model relative to the three-parameter moments model, and find no evidence to reject the three-parameter moments model relative to the general model.
    }
  \label{fig_elr}
\end{figure}

As a concrete example, we conduct a simulated empirical likelihood ratio test to check for deviations from the basic model of \fr{} benchmarking.
We consider a model where the SPAM error is fixed at $\theta_0 = 3\times10^{-2}$ and for each \shot{} the step error $\theta_1$ is drawn independently from a Gaussian distribution with mean $1\times10^{-4}$ and standard deviation $2.5\times 10^{-5}$.
To choose an experiment design for the simulated experiment, we use the four-parameter moments model and perform the optimization described in Section~\ref{sec_opt}.
For this optimization we choose the reference point to match the moments of the actual Gaussian distribution of the step error, and we minimize the standard deviation of the parameter $\theta_2$ as a proxy for maximizing the statistical power to reject the basic model.
We choose $\theta_2$ as a proxy because $\theta_3$ is zero for the chosen distribution of step errors.
In the experiment optimized for $\theta_2$, the standard deviation of the step error $\theta_1$ is $2.76\times10^{-6}$. 
If the experiment were instead optimized to minimize the standard deviation of $\theta_1$, the optimal experiment in that case would have a standard deviation for $\theta_1$ of $1.96\times10^{-6}$. 
This illustrates the fact that the decision to optimize the experiment to maximize statistical power to reject the basic model has a relatively small effect on the performance of inferring the step error.
The optimized experiment is constrained so that the total run time is $30$ minutes, assuming that each step takes $10^{-5}$ s and state preparation and measurement takes $10^{-3}$ s.
Once the experiment design has been chosen, we simulate a dataset for this experiment by drawing a step error $\theta_1$ independently for each \shot{}.
Once a value of $\theta_1$ has been drawn, we then draw `success' or `failure' with the corresponding probability obtained from the basic model for the drawn value of $\theta_1$ and the particular sequence length in question.
With the simulated dataset we then perform a bootstrapped empirical likelihood ratio test between the basic model, three-parameter moments model, and the general model, which are nested models.
For each choice of inner model and outer model we follow the procedure outlined in the previous paragraph, and the results are shown in Fig.~\ref{fig_elr}.
According to the distribution of bootstrapped likelihood ratios, the p-value to reject the basic model relative to the three-parameter moments model is $4.9\%$.
 No significant deviation from the three-parameter moments model relative to the general model was detected (p-value of 59.2\%).
 These results match the intuition that the Gaussian fluctuation in the step error is detectable via the second moment, and that the fourth and higher moments can be safely neglected in this scenario.

\section{Experimental Implementation}
\label{sec_exp_implementation}

To provide a concrete comparison between non-fully-randomized benchmarking and optimized \fr{} benchmarking, we designed and implemented three randomized benchmarking experiments.
To realize these experiments we perform single qubit
rotations on a $^{25}$Mg$^+$ ion in a microfabricated surface-electrode ion trap, in
the apparatus described in Refs.~\cite{burd_quantum_2019, srinivas_high_fidelity_2021}.
We use the states $\ket{F=3, m_F=1}$ (logical $\ket{1}$) and $\ket{F=2, m_F=1}$ (logical $\ket{0}$)
in the $^2S_{1/2}$ ground-state hyperfine manifold to realize a qubit.
The qubit transition frequency of $\omega = 2 \pi \times 1686$ MHz is first-order
insensitive to the magnetic field at $B \approx$ 213 G,
mitigating against errors caused by fluctuations in the total magnetic field.
Qubit rotations around X and Y are implemented with microwave magnetic fields applied at
the transition frequency with differing phase, while Z rotations are
implemented by adding a phase offset to the microwave control signal for subsequent rotations.
The qubit is prepared with optical pumping followed by microwave pulses to transfer
population to the $\ket{1}$ state.
Qubit readout is accomplished by applying a laser resonant with the $^2S_{1/2}$ to
$^2P_{3/2}$ cycling transition and detecting state-dependent ion fluorescence as in Ref.~\cite{srinivas_high_fidelity_2021} (SM).
Full randomization is achieved by choosing gates on the fly in real time with a
pseudorandom number generator (PRNG) \cite{oneill_pcg2014}
running on the same FPGA (field programmable gate array) that is used to generate the gate pulses applied to the ion.
The ideal stabilizer (assuming no errors) is stored and concurrently
updated on the FPGA as new gates are chosen, such that when the required number
of random gates have been applied the stabilizer can be used to return the
qubit to the measurement basis and indicate the expected measurement outcome.
The on-the-fly calculation process for sequences can also be configured to enable intentional repetition of random gate sequences.

The three experiments in the comparison are as follows.
First, we constructed an experiment where $10$ sequence lengths were set uniformly in the range from $5$ to $1/x_0$ where $x_0 = 2\times 10^{-5}$ is the best guess for the step error prior to the experiment.
At each sequence length we drew $24$ random sequences and repeated each of them $24$ times.
This experiment took roughly $53.5$ minutes of total time.
Then, we repeated the same experiment but fully randomized the sequences, so at each sequence length a total of $24 \times 24 = 576$ random sequences were drawn and run once.
The time to run the experiment is unaffected by fully randomizing, so this experiment also took $53.5$ minutes of total time.
Finally, we designed an optimized, \fr{} experiment using the methods in Section \ref{sec_opt}.
The reference point for the optimization has a SPAM parameter of $3 \times 10^{-2}$ and a step error parameter of $2 \times 10^{-5}$, which were determined prior to the experiment by calibrations and shorter RB experiments. 
The optimization minimizes the standard deviation of the step error according to the four-parameter moments model.
The total time of the optimized experiment was constrained to match the total time of the non-optimized experiments.
For experimental simplicity we rounded the number of trials at each sequence length to a multiple of four, so that each experiment could be divided into four equal blocks. 
Within the first block, the order of experimental trials is randomly chosen and then the same order of trials is repeated for the remaining three blocks.
Rounding the number of trials at each sequence length to a multiple of four had a negligible effect on the wall-clock time and anticipated standard deviations.

\begin{figure}[h!]
  \captionsetup[subfloat]{labelfont=bf, farskip=0mm}
  \subfloat[Uniform design, repeated sequences] {%
  \includegraphics[width=.5\textwidth]{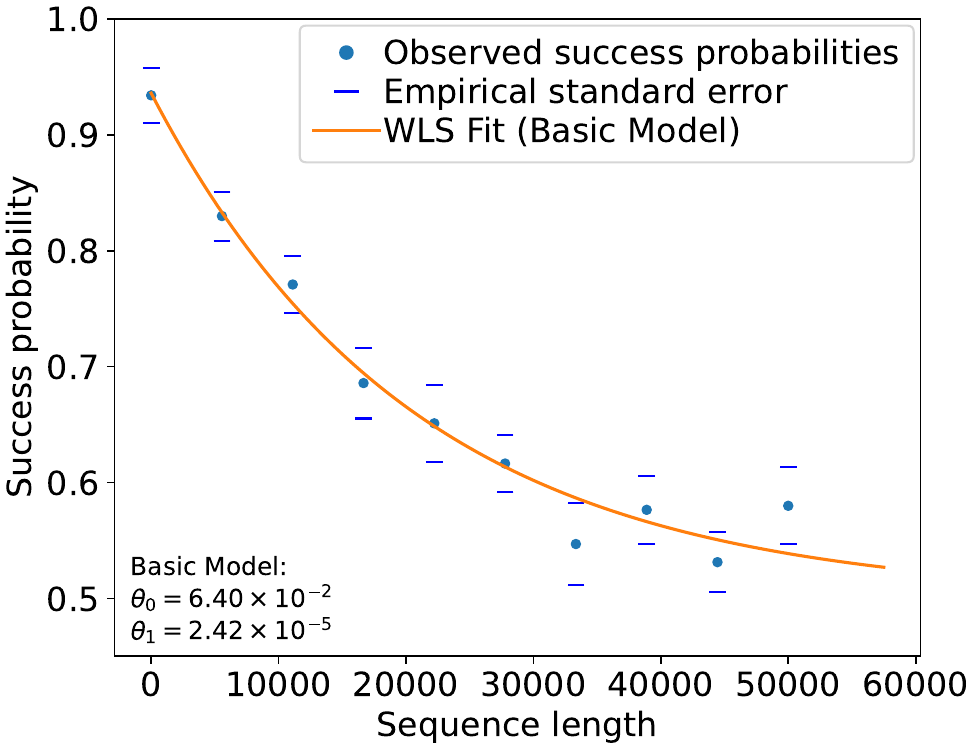}%
  }
  \subfloat[Uniform design, \fr{}] {%
    \includegraphics[width=.5\textwidth]{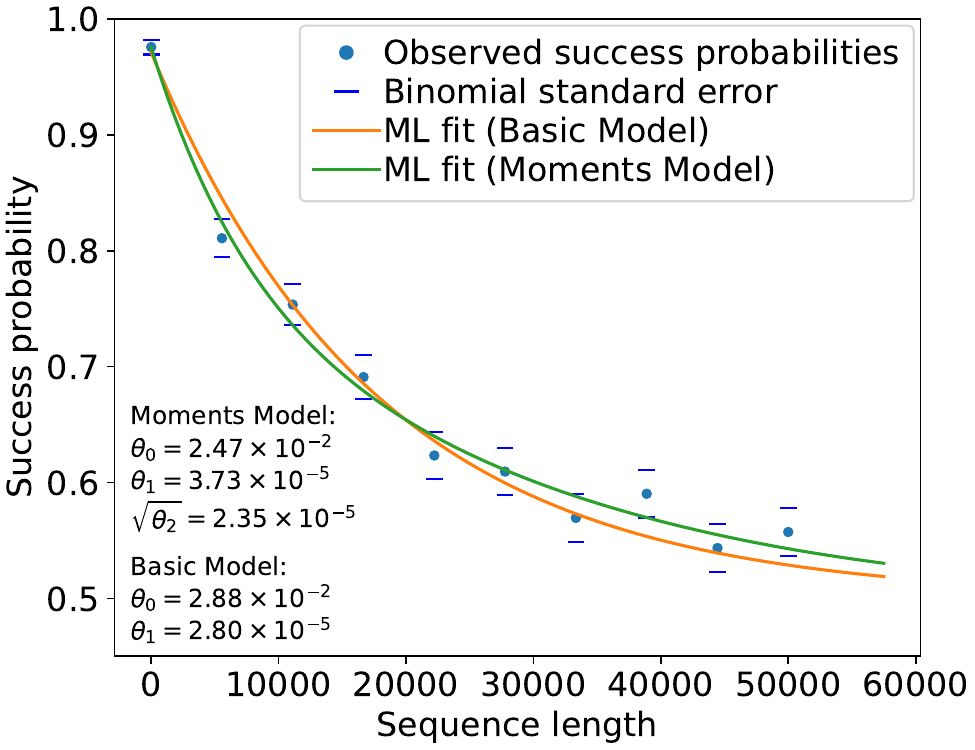}%
  }

  \subfloat[Optimized design, \fr{}] {%
    \includegraphics[width=.5\textwidth]{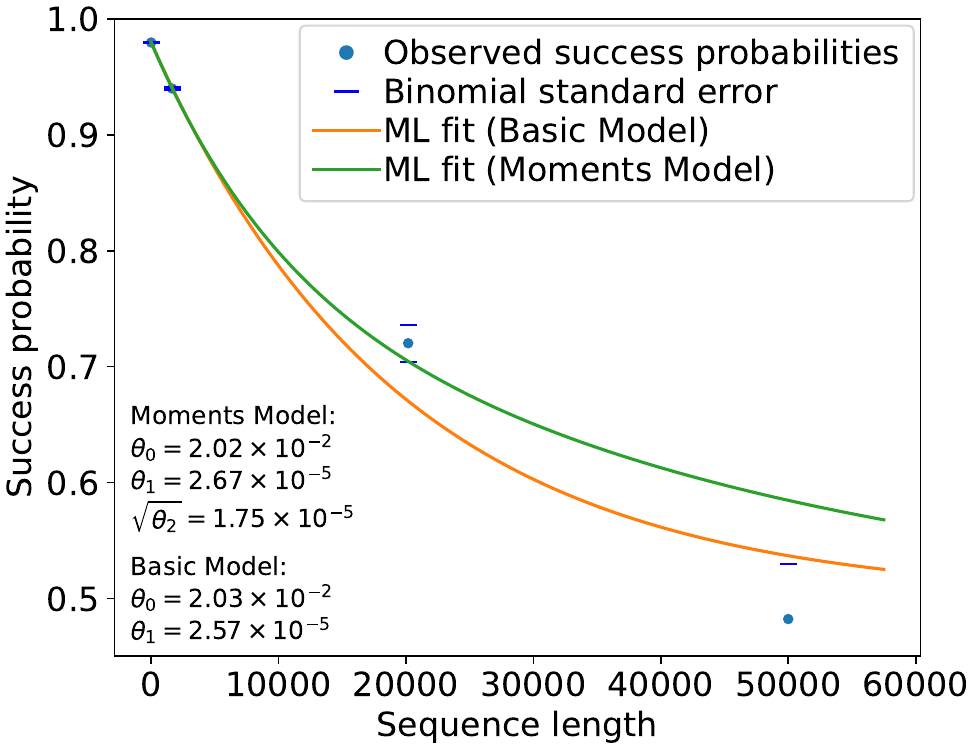}
  }
  \caption{The observed decays in success probability for each of the three randomized benchmarking experiments run at NIST comparing non-\fr{} benchmarking to optimized \fr{} benchmarking. The experiment in plot (a) has sequence lengths chosen uniformly in the range $[5,5\times 10^{4}]$ and for each sequence length $24$ random sequences are drawn and run $24$ times each. The orange trace is the best fit to the basic model, obtained by a weighted least squares fit to the observed success probabilities, and the best fit parameters are shown inset in the lower left. The weights in the fit are the squared inverses of empirical standard errors of the observed success probabilities at each sequence length. These empirical standard errors are shown with the blue tickmarks. The experiment in plot (b) has the same sequence lengths as the first experiment, but is fully randomized so at each sequence length $24\times24$ random sequences are drawn and run once each. The experiment in plot (c) is designed according to the optimization routine in Section~\ref{sec_opt}, where the total experiment time is constrained to match the total experiment time of each of the previous two experiments. In plots (b,c) the orange and green traces are the maximum likelihood fits to the basic model and the three-parameter moments model respectively, and the maximum likelihood parameters are shown inset in the lower left. The blue ticks show the binomial standard errors of the observed success probabilities. 
  }
  \label{fig_main_plots}
\end{figure}

\begin{figure}[h!]
  \captionsetup[subfloat]{labelfont=bf, farskip=-3mm}
  \subfloat[]{\includegraphics[width=.5\textwidth]{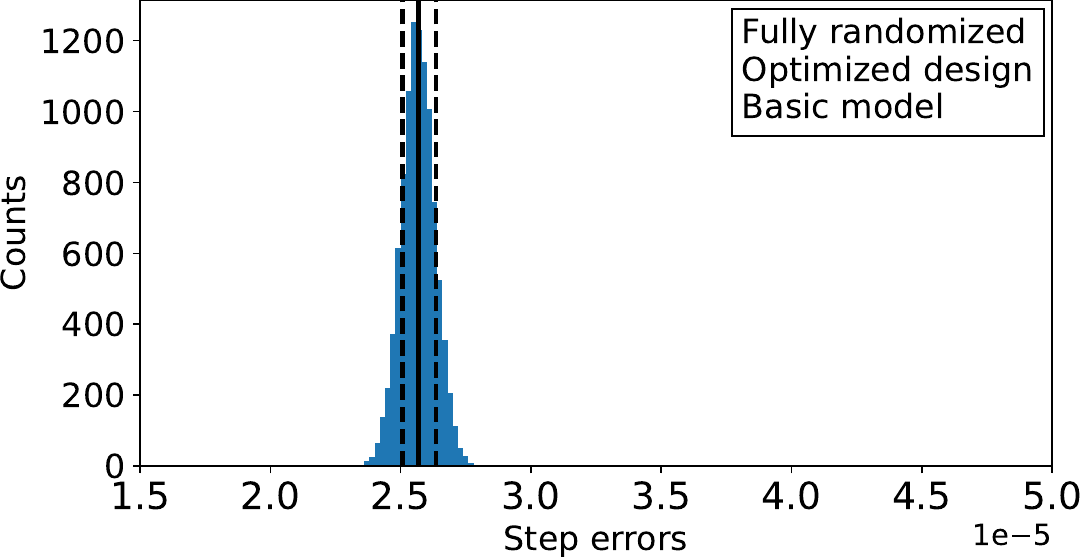}}%
  \subfloat[]{\includegraphics[width=.5\textwidth]{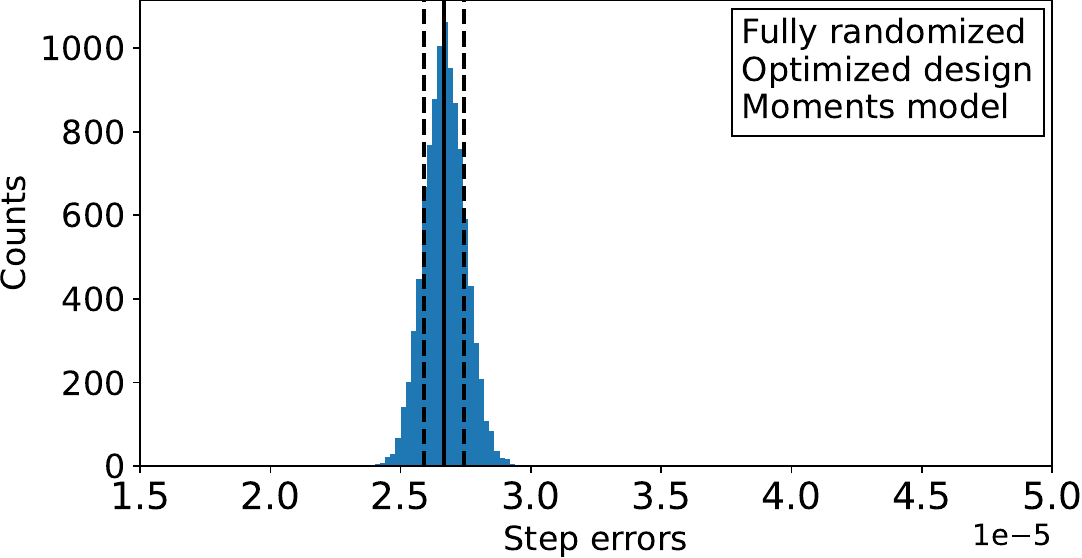}}%
 
  \subfloat[]{\includegraphics[width=.5\textwidth]{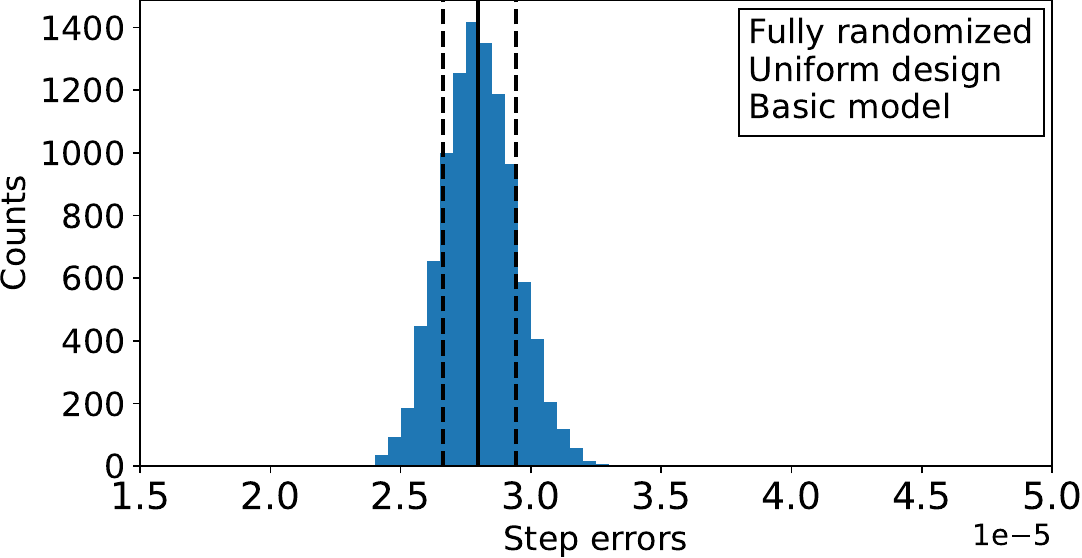}}%
  \subfloat[]{\includegraphics[width=.5\textwidth]{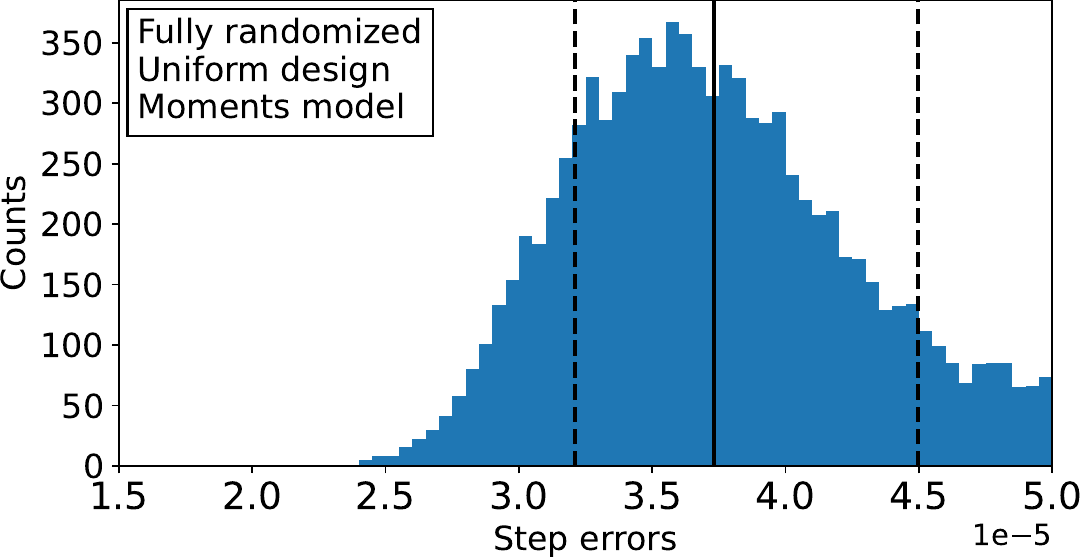}}%

  \subfloat[]{\includegraphics[width=.5\textwidth]{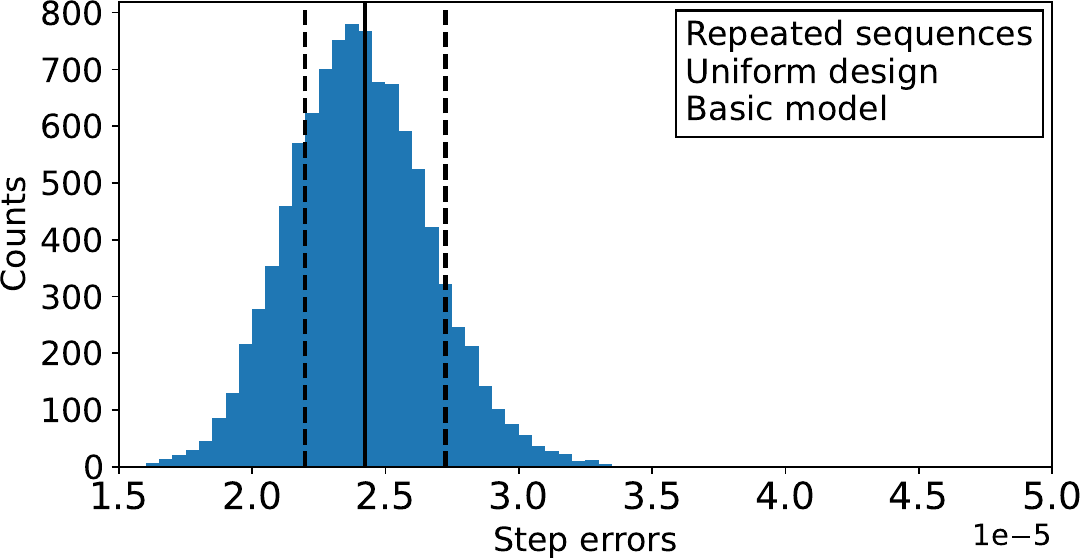}}%
  \hspace*{.5\textwidth}
  \caption{The bootstrap distributions obtained during the analysis of the three experiments run at NIST during our comparison between non-fully-randomized benchmarking and optimized \fr{} benchmarking. 
  The plots in the left column correspond to the analysis according to the basic model and the plots in the right column correspond to the analysis according to the three-parameter moments model. 
  The plots in the first row are for the optimized, \fr{} experiment, the plots in the second row are for the uniform, \fr{} experiment, and the plot in the third row is for the uniform experiment with repeated sequences. 
  We do not include the plot for the uniform experiment with repeated sequences analyzed according to the moments model because performing a weighted least squares fit to the moments model is not a standard technique in randomized benchmarking. 
  For all the plots, the solid black line indicates the step error parameter of the best fit to the original data and the dashed black lines denote the $68\%$ confidence interval obtained with bias-corrected bootstrapping. 
The best fit step errors for the plots are (a) $\dataMainBasicOptFR$, (b) $\dataMainMomentsOptFR$, (c) $\dataMainBasicOrigFR$, (d) $\dataMainMomentsOrigFR$, (e) $\dataMainBasicOrig$.
  }
  \label{fig_main_bts_comp}
\end{figure}

To analyze the randomized benchmarking experiment with repeated sequences, we ran a weighted least squares fit to the basic model. The weights in the fit are the squared inverses of the empirical standard errors of the success probabilities at each sequence length.
The empirical standard errors are obtained by computing the empirical standard deviation of the estimated success probabilities of the random sequences and dividing by the square root of the number of sequence repetitions. For further information about weighted least square fits in randomized benchmarking we refer to Ref.~\cite{meier_thesis}.
To analyze the \fr{} experiments we perform the maximum likelihood inference that we outline in Section~\ref{sec_statistics}. We perform this maximum likelihood analysis for both the basic model and the moments model with three total parameters.
The results for all three experiments are shown in Fig.~\ref{fig_main_plots}.
To obtain confidence intervals on the step error for the various experiments, we perform bias-corrected parametric bootstrapping with $10,000$ bootstrap samples, as described in Section~\ref{sec_statistics} and in Ref.~\cite{efron_bootstrap}.
For the first experiment, which has intentionally repeated sequences, the bootstrap samples are obtained by following the procedure in Ref.~\cite{meier_thesis}. To summarize, first we resample the list of sequences with replacement, and then for each sequence we binomially resample the success and failure counts. Then, for each bootstrapped dataset the step error is estimated with a weighted least squares fit to the model. 
For the second and third experiments, which are \fr{}, the bootstrap samples are obtained by parametrically resampling according to the parameters obtained from the maximum likelihood analysis on the original data.
The bootstrap histograms, point estimates, and $68\%$ bootstrapped confidence intervals are shown in Fig.~\ref{fig_main_bts_comp}, where we run the analysis according to both the basic model and the three-parameter moments model. For the uniform design with repeated sequences, we report a step error of $\dataMainBasicOrig$ when analyzing according to the basic model. 
For the optimized \fr{} experiment we report a corresponding step error of $\dataMainBasicOptFR$, which has a confidence interval that is roughly four times smaller.
To test the basic model of the optimized, \fr{} experiment, we performed the empirical likelihood ratio test described in Section 7. 
The results are shown in Fig. 5. We found a p-value of $0.3\%$ to reject the basic model. This shows evidence of deviation from an exponential decay, which we interpret as evidence of non-Markovian or time-dependent behavior.

\begin{figure}
  \includegraphics[width=.5\textwidth]{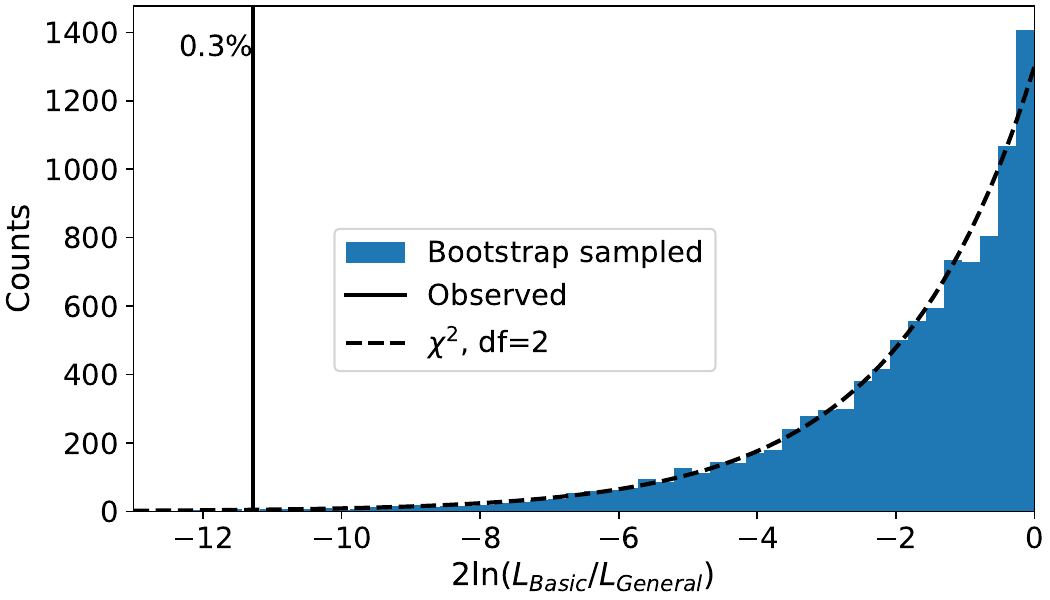}%
  \caption{
    Results of an empirical likelihood ratio test for the optimized, \fr{} experiment. 
  We perform the procedure described in Section~\ref{sec_statistics}, with the basic model as the inner model and the general model as the outer model.
  We find a p-value to reject the basic model of $0.3\%$.
  This shows evidence of deviation from an exponential decay, which we interpret as evidence of non-Markovian or time-dependent behavior.}
  \label{fig_main_elr}
\end{figure}

\begin{figure}
  \subfloat[Uniform design, repeated sequences] {%
  \includegraphics[width=.5\textwidth]{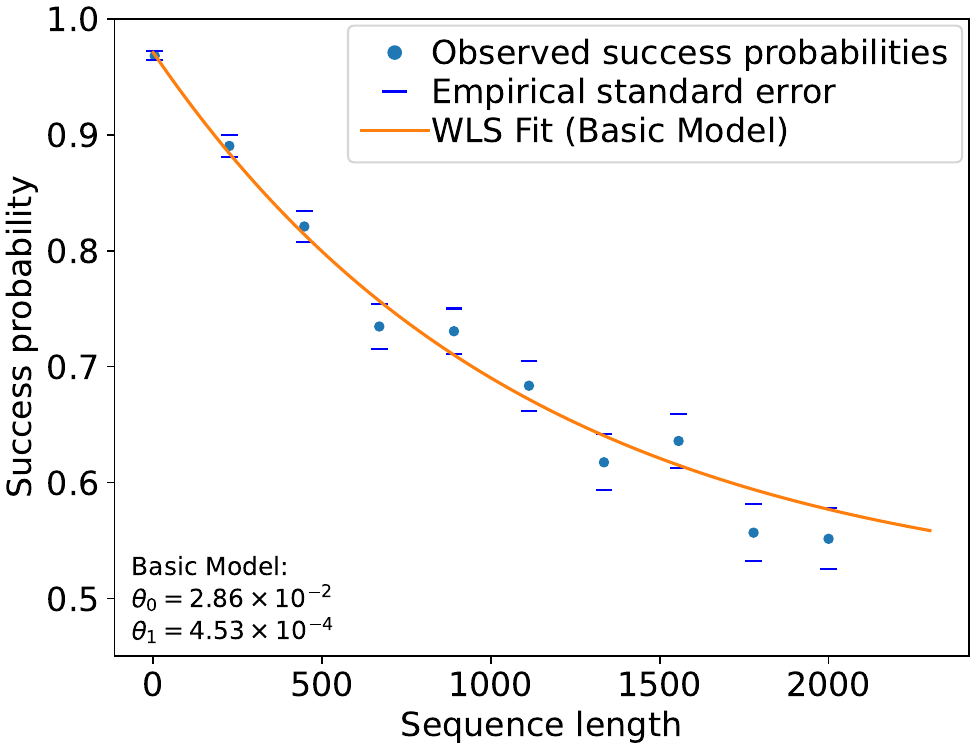}%
  }
  \subfloat[Uniform design, \fr{}] {%
    \includegraphics[width=.5\textwidth]{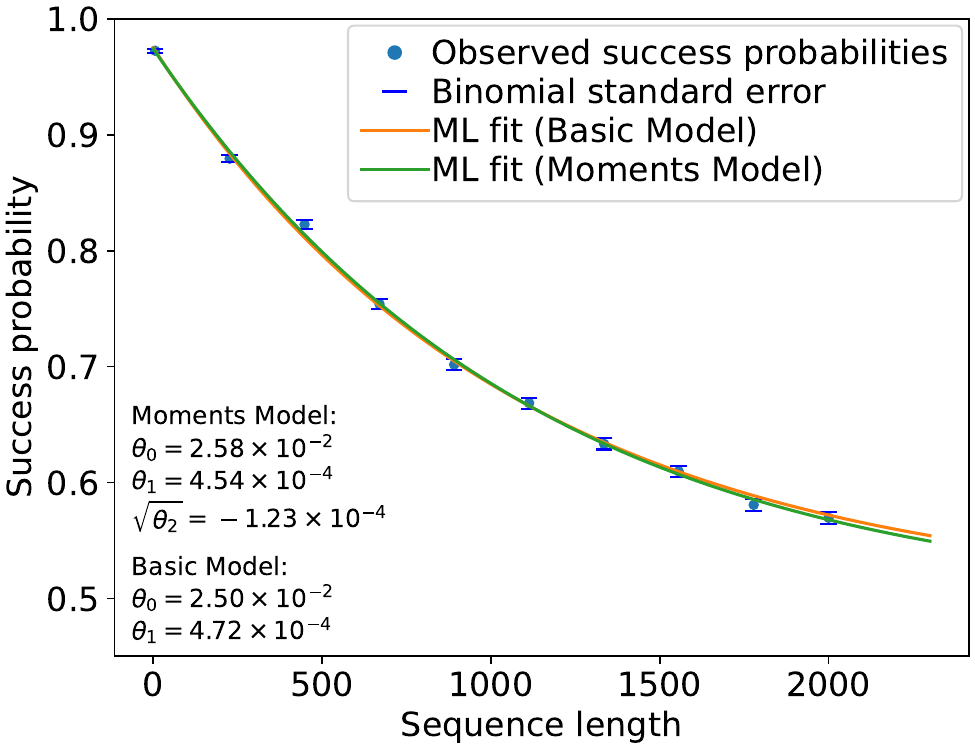}%
  }

  \subfloat[Optimized design, \fr{}] {%
    \includegraphics[width=.5\textwidth]{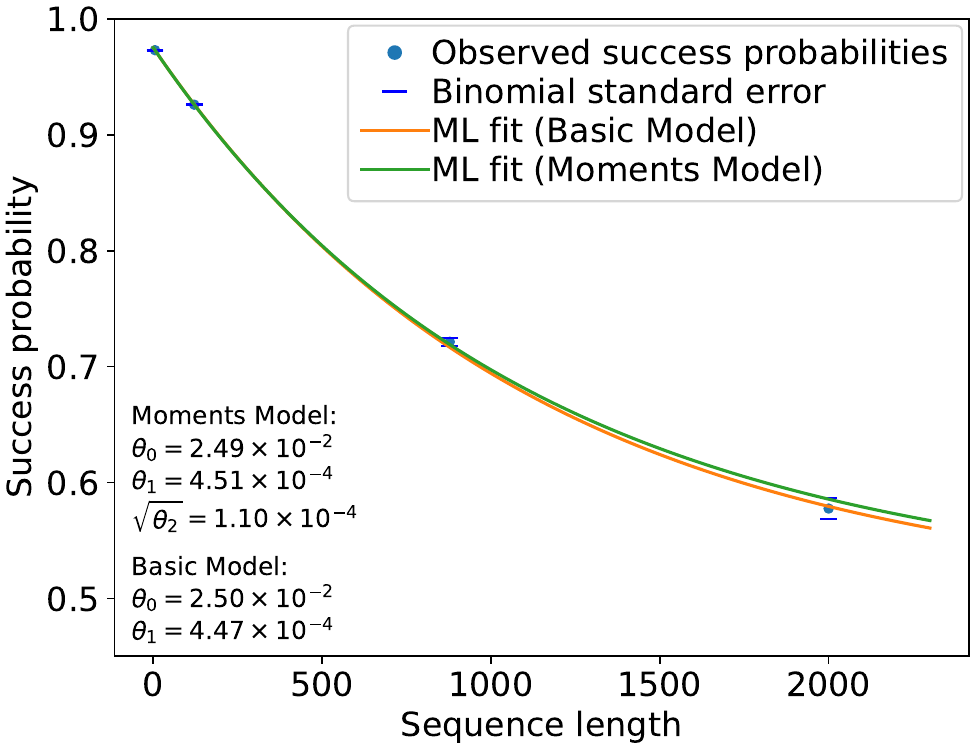}
  }
  \caption{The observed decays in success probability for each of the three experiments where we intentionally introduced coherent errors. The experiment in plot (a) has sequence lengths chosen uniformly in the range [5,2000] and for each sequence length $100$ random sequences are drawn and run $100$ times each. The experiment in plot (b) has the same sequence lengths as the first experiment, but is fully randomized so at each sequence length $100\times100$ random sequences are drawn and run once each. Each of the three experiments takes the same total time of approximately $40$ minutes. All other aspects of the plots are the same as in Fig.~\ref{fig_main_plots}.}
  \label{fig_coherent_plots}
\end{figure}

After completing these three experiments, we intentionally introduced a unitary error by miscalibrating the gates in the 2-design and repeated the same comparison between non-fully-randomized benchmarking and optimized \fr{} benchmarking.
The size of the miscalibration was chosen to give a step error of approximately $5\times10^{-4}$. We followed the same procedure that we used previously to construct three randomized benchmarking experiments.
For the first experiment we chose $10$ sequence lengths uniformly in the range $[5,2000]$, where the maximum sequence length again corresponds to $1/x_0$. 
At each sequence length we drew 100 random sequences and repeated each of them 100 times. 
This experiment took roughly $40$ minutes of total time.
Second, we repeated the same experiment but fully randomized the sequences so at each sequence length $100\times100$ random sequences were drawn and run once. 
Third, we performed an optimized \fr{} experiment that took the same wall-clock time, where the optimization was again done to maximize statistical power to infer the step error using the four-parameter moments model. 
The results of this analysis are reported in Fig.~\ref{fig_coherent_plots} and the bootstrap distributions are reported in Fig.~\ref{fig_coherent_bts_comp}. 
We again observe a confidence interval for the optimized \fr{} experiment that is roughly $4$ times smaller than that of the uniform experiment with repeated sequences. 
We also run the same empirical likelihood ratio test between the basic model and the general model. 
The results are shown in Fig.~\ref{fig_coherent_elr} and we observe no significant deviation from the basic model. 
\newcommand{\designMainBasicOrigfr}{9.37\times 10^{-7}}
\newcommand{\designMainBasicOpt}{6.49\times 10^{-7}}
\newcommand{\designMainMomentsOrigfr}{2.18\times 10^{-6}}
\newcommand{\designMainMomentsOpt}{7.51\times 10^{-7}}
\newcommand{\designMainTotalTime}{3219.12}
\newcommand{\designMainRefStep}{2.0\times 10^{-5}}

\newcommand{\designCoherentBasicOrigfr}{5.68\times 10^{-6}}
\newcommand{\designCoherentBasicOpt}{4.94\times 10^{-6}}
\newcommand{\designCoherentMomentsOrigfr}{1.33\times 10^{-5}}
\newcommand{\designCoherentMomentsOpt}{6.41\times 10^{-6}}
\newcommand{\designCoherentTotalTime}{2368.91}
\newcommand{\designCoherentRefStep}{5.0\times 10^{-4}}

\newcommand{\dataCoherentBasicOrig}{4.53^{+0.26}_{-0.18} \times 10^{-4}}
\newcommand{\dataCoherentBasicOrigFR}{4.72^{+0.05}_{-0.05} \times 10^{-4}}
\newcommand{\dataCoherentBasicOptFR}{4.47^{+0.05}_{-0.04} \times 10^{-4}}
\newcommand{\dataCoherentMomentsOrig}{4.05^{+0.50}_{-0.35} \times 10^{-4}}
\newcommand{\dataCoherentMomentsOrigFR}{4.54^{+0.12}_{-0.11} \times 10^{-4}}
\newcommand{\dataCoherentMomentsOptFR}{4.51^{+0.06}_{-0.06} \times 10^{-4}}

\begin{figure}[ht]
  \captionsetup[subfloat]{labelfont=bf, farskip=0pt}
  \subfloat[]{\includegraphics[width=.5\textwidth]{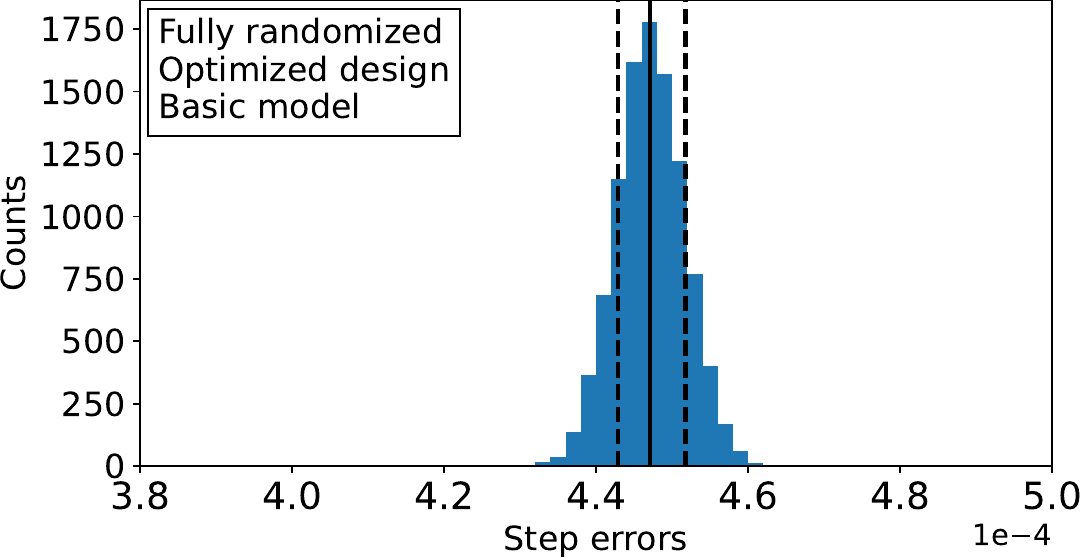}}%
  \subfloat[]{\includegraphics[width=.5\textwidth]{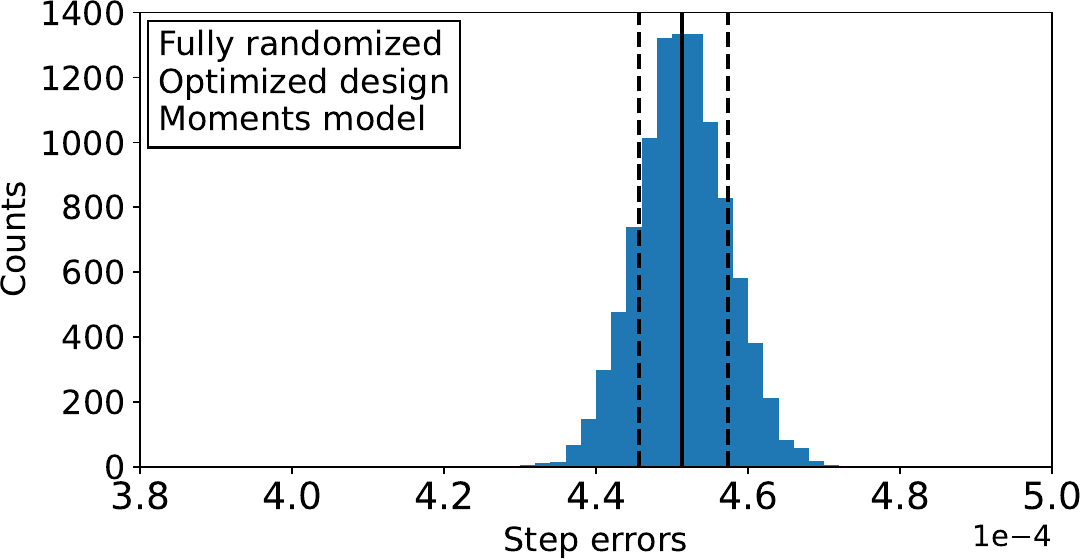}}%
 
  \subfloat[]{\includegraphics[width=.5\textwidth]{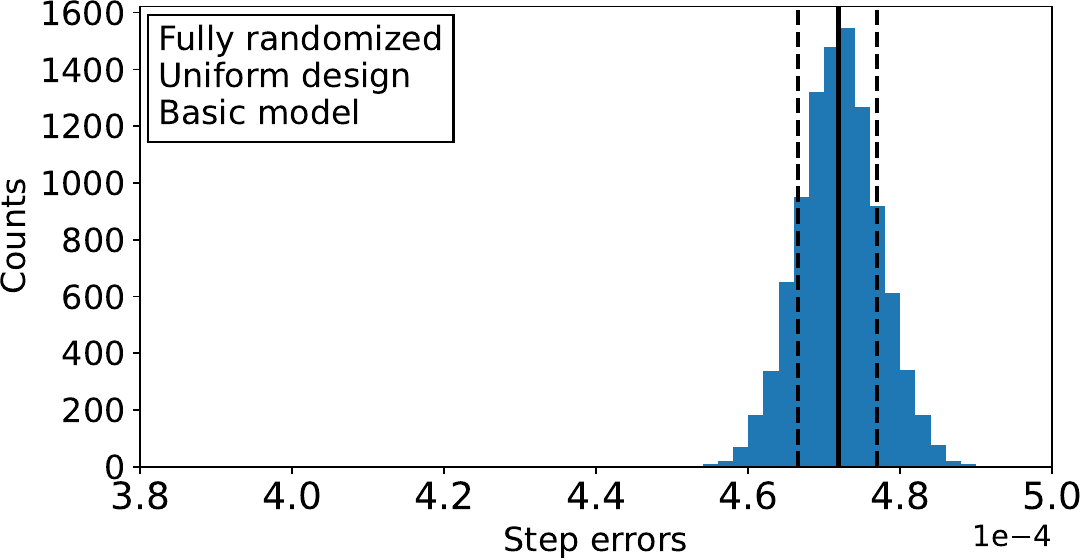}}%
  \subfloat[]{\includegraphics[width=.5\textwidth]{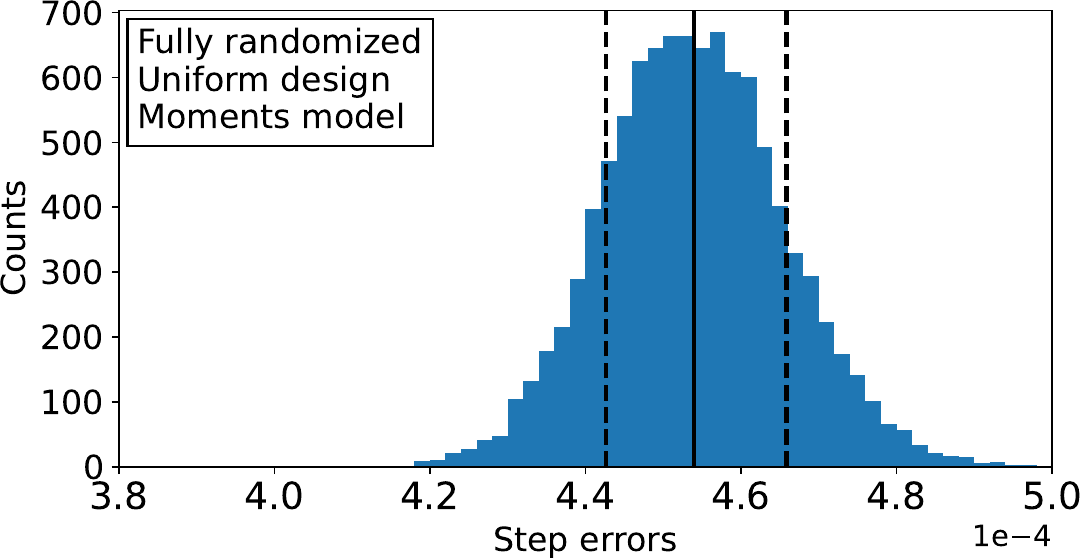}}%

  \subfloat[]{\includegraphics[width=.5\textwidth]{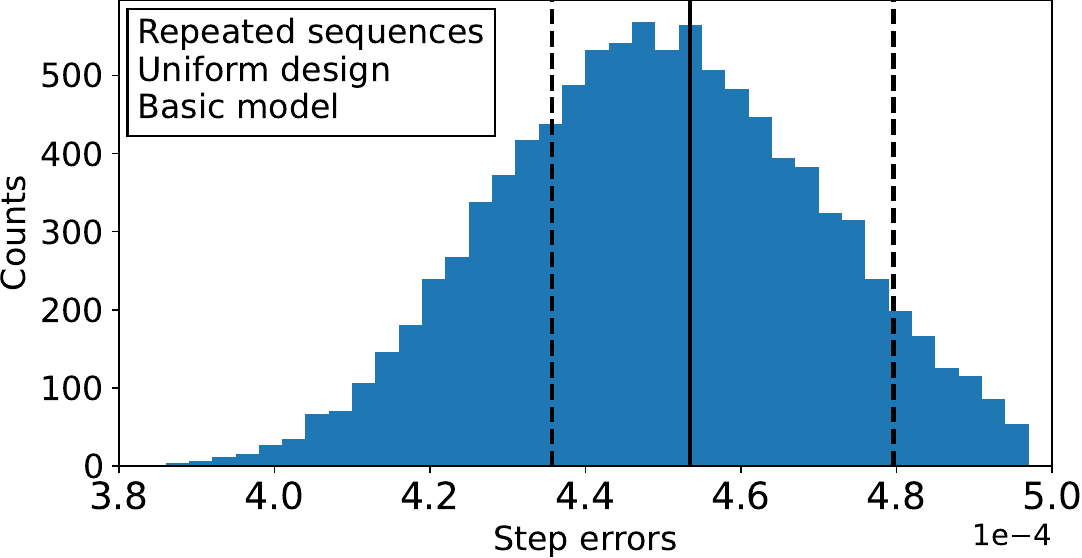}}%
  \hspace*{.5\textwidth}
  \caption{The bootstrap distributions obtained during the analysis of the three experiments we ran during the comparison where we intentionally introduced coherent errors. All aspects of the plots are the same as in Fig.~\ref{fig_main_bts_comp}. The best fit step errors for the plots are (a) $\dataCoherentBasicOptFR$, (b) $\dataCoherentMomentsOptFR$, (c) $\dataCoherentBasicOrigFR$, (d) $\dataCoherentMomentsOrigFR$, (e) $\dataCoherentBasicOrig$.}
  \label{fig_coherent_bts_comp}
\end{figure}

\begin{figure}
  \subfloat[Basic model vs Moments model] {%
  \includegraphics[width=.5\textwidth]{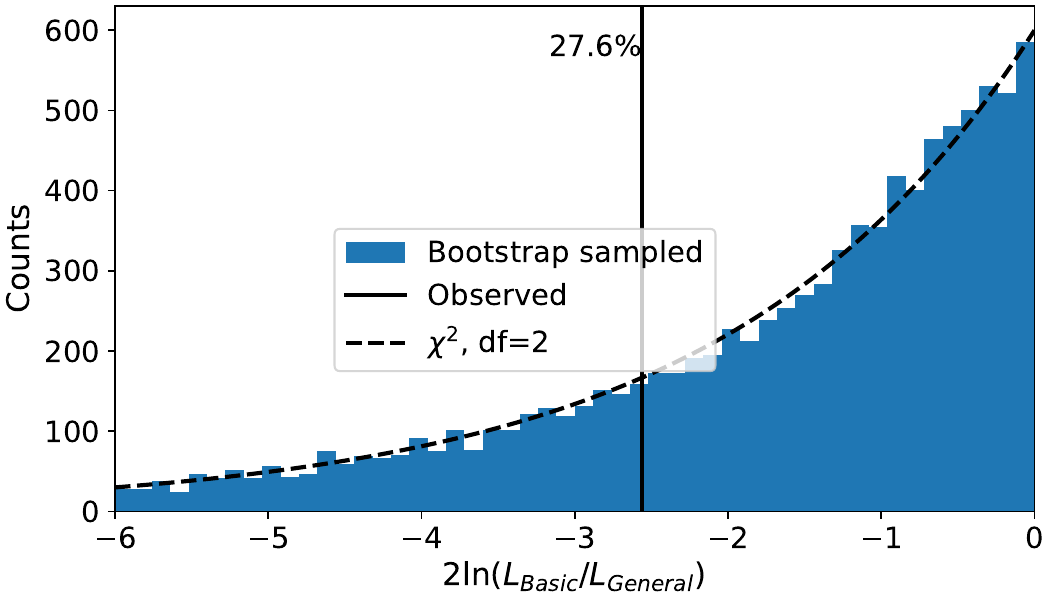}%
  }
  \caption{Results of an empirical likelihood ratio test for the optimized, \fr{} experiment that we ran as part of the comparison where we intentionally introduced coherent errors. The p-value to reject the basic model is $27.6\%$. Thus we find no significant evidence to reject the basic model. All other aspects of the plots are the same as in Fig.~\ref{fig_main_elr}.}
  \label{fig_coherent_elr}
\end{figure}
\section{Conclusion}
In this work we study \fr{} benchmarking, where a new random sequence is drawn independently for each \shot{}.
We analyze the concrete advantages of \fr{} benchmarking, which include smaller error bars on the inferred step error, maximum likelihood analysis without heuristics, straightforward optimization
of the sequence lengths, insensitivity to drifts in SPAM throughout the experiment, and the ability to model and measure behaviors
such beyond the basic randomized
benchmarking model usually assumed, such as gate-position-dependent errors or time-drifting errors.
Furthermore, we provide a general formulation of statistical models for \fr{} benchmarking and give a procedure to optimize the design of the experiment to minimize the uncertainty of inference of a particular model parameter, typically the step error.
This optimization can be done for an arbitrary statistical model that can be linearized around a reference point, and takes into account the actual wall-clock time of running a random sequence of each possible length.
For experiments that are not \fr{}, we analyze the dependence of the uncertainty on the number of times that each sequence is repeated and show concrete advantages from fully randomizing.
We also discuss the moments model of \fr{} benchmarking and show that it is a general model of time-dependent errors when constraints on the moments parameters are removed. 
We show how an empirical likelihood ratio test can be used to possibly distinguish the basic model of a single exponential decay from more general models.
In fact, one key advantage of \fr{} benchmarking is that likelihood ratio tests become straightforward to implement and interpret, largely because maximum-likelihood inference can be used to analyze a \fr{} experiment.
Finally, we implement \fr{} benchmarking on a trapped ion qubit at NIST and run experiments that allows us to compare optimized, \fr{} benchmarking to randomized benchmarking with uniform sequence lengths and intentionally repeated sequences. We find substantial reductions in the uncertainty in the estimated step error as a result of fully randomizing.

\appendix
\section{Computing optimal linear estimators for a given experiment design}
\label{sec_opt_le}
If the experiment design is fixed, the coefficients $\listp{C_n}$ of the optimal linear estimator for a parameter $\theta_{i_0}$ at the reference point $\paramrefpoint$ can be computed as follows.
In the notation of Section~\ref{sec_opt}, the goal is to minimize the variance in Eq.~\ref{eq_lin_est_var}, which is $v = \sum_n \frac{C_n^2v_n}{w_n}$, subject to the linear constraints $\sum_n C_n L_{ni} = \delta_{ii_0}$.
This is a quadratic program with linear constraints and can be written in matrix notation as a minimization of
 $\mathbf{c}^\top Q \mathbf{c}$ subject to $E\mathbf{c} = \mathbf{d}$, where $\mathbf{c}$ is the list of coefficients $\listp{C_n}$ in vector form, $Q$ is a diagonal matrix with diagonal elements $Q_{nn} = \frac{v_n}{w_n}$, the constraint matrix $E$ has elements $E_{in} = L_{ni}$, and $d_i = \delta_{ii_0}$.
 The solution for $\mathbf{c}$ can be obtained by solving
\begin{equation}
  \mqty[Q & E^\top \\ E & 0]\mqty[\mathbf{c} \\ \mathbf{\lambda}]=\mqty[0 \\ \mathbf{d}]
\end{equation}
where $\mathbf{\lambda}$ is a vector of Lagrange multipliers \cite{quadratic_programming}. This can be achieved by using the standard formula for the inverse of a block matrix \cite{quadratic_programming, block_matrix_inverse}, and the solution for $\mathbf{c}$ is
\begin{equation}
  \mathbf{c} = Q^{-1}E^\top(EQ^{-1}E^\top)^{-1}\mathbf{d}.
\end{equation}
As a result, the $i$th column of the matrix $M = Q^{-1}E^\top(EQ^{-1}E^\top)^{-1}$ has the coefficients of the optimal linear estimator for the $i$th parameter $\theta_i$.
\section{Interpretation of design optimization in the context of Fisher information}
\label{sec_fisher_interpretation}
For a given experiment design and reference point $\paramrefpoint$, we show that the covariance matrix $V$ of the optimal linear estimators obtained in Appendix~\ref{sec_opt_le} is equal to the inverse of the Fisher information matrix.
This is well established in the literature on experiment design and Fisher information \cite{nielsen_cr,nielsen_info_geometry,fedorov_exp_design,pukelsheim_exp_design}, and for convenience we provide a derivation here. 
As we show in Appendix~\ref{sec_opt_le}, the $i$th column of the matrix $M = Q^{-1}E^\top(EQ^{-1}E^\top)^{-1}$ has the coefficients of the optimal linear estimator for the $i$th parameter $\theta_i$.
Therefore, the covariance matrix $V$ of these linear estimators satisfies $V = M^\top Q M$, which evaluates to
\begin{equation}
  V = M^\top Q M = (EQ^{-1}E^\top)^{-1\top} E Q^{-1} Q Q^{-1} E^\top (E Q^{-1} E^\top)^{-1}.
\end{equation}
The matrix $Q$ is diagonal, so we have $(EQ^{-1}E^\top)^{-1\top} = (EQ^{-1}E^\top)^{-1}$,
and this simplifies to
\begin{equation}
  \label{eq_lin_est_cov}
  V = (EQ^{-1}E^\top)^{-1}.
\end{equation}

The Fisher information matrix for a single \shot{} of sequence length $n$ can be obtained according to the standard formula \cite{nielsen_cr}
\begin{equation}
  F_{ii^\prime }(n) = \bigg<\frac{\partial}{\partial\theta_i}\log{p}\frac{\partial}{\partial\theta_{i^\prime} }\log{p}\bigg>_{\paramrefpoint},
\end{equation}
where the expectation value is taken over the two measurement outcomes `success' and `failure', and $p$ is the likelihood of getting a particular outcome.
The subscript $\paramrefpoint$ indicates that the formula is evaluated at the reference point $\paramrefpoint$.
Evaluating this for the two-outcome measurement for a single \shot{} of sequence length $n$ gives
\begin{align}
  F_{ii^\prime }(n) = & \left.P_{\boldsymbol\theta}(n)\frac{\partial}{\partial\theta_i}\blog{P_{\boldsymbol\theta}(n)}\frac{\partial}{\partial\theta_{i^\prime} }\blog{P_{\boldsymbol\theta}(n)}\right|_{\paramrefpoint} + \nonumber \\
  & \left.\left[1-P_{\boldsymbol\theta}(n)\right]\frac{\partial}{\partial\theta_i}\blog{1-P_{\boldsymbol\theta}(n)}\frac{\partial}{\partial\theta_{i^\prime} }\blog{1-P_{\boldsymbol\theta}(n)}\right|_{\paramrefpoint}.
\end{align}
Using the fact that $\left.E_{ni} = \frac{\partial}{\partial\theta_i}P_{\boldsymbol\theta}(n)\right|_{\paramrefpoint}$, this simplifies to
\begin{equation}
  F_{ii^\prime }(n) = \frac{E_{ni}E_{ni^\prime }}{P_{\boldsymbol\theta^{(0)}}(n)(1-P_{\boldsymbol\theta^{(0)}}(n))}.
\end{equation}
Weighting by the number of \shots{} $w_n$ at sequence length $n$ and summing over $n$ gives a total Fisher information matrix of
\begin{equation}
  F_{ii^\prime} = \sum_n \frac{E_{ni}w_nE_{i^\prime n}}{P_{\boldsymbol\theta^{(0)}}(n)(1-P_{\boldsymbol\theta^{(0)}}(n))}.
\end{equation}
Using the fact that, as in Section~\ref{sec_opt}, the matrix $Q$ is diagonal with entries $Q_{nn} = w_n/v_n$ with $v_n = [P_{\boldsymbol\theta^{(0)}}(n)(1-P_{\boldsymbol\theta^{(0)}}(n))]$, this simplifies to $F = E Q^{-1} E^\top$.
Therefore, comparison to Eq.~\ref{eq_lin_est_cov} shows that the covariance matrix of the optimal linear estimators $V$ is equal to the inverse of the Fisher information matrix $F$.
In this sense, the optimization procedure described in Section~\ref{sec_opt} is Fisher-optimal.
\section{Details of variance analysis of \fr{} benchmarking}
\label{app_excess_variance}
  Here we verify the fact that
\begin{equation}
  \int d\psi_H f_\psi^2 = \frac{2}{D(D+1)},
\end{equation}
where $d\psi_H$ denotes the Haar measure over pure states, and $f_\psi$ is the fidelity of the random pure state $\ket\psi$ with the target state $\ket\chi$.
The Haar-random pure state $\ket\psi$ can be expressed as $U\ket{\chi}$ for a Haar-random unitary $U$, so this integral can be written as
\begin{equation}
  \int d\psi_H f_\psi^2 = \int dU_H \btr{(U \otimes U)\ketbra{\chi}^{\otimes 2}(U^\dagger \otimes U^\dagger) \ketbra{\chi}^{\otimes 2}}.
\end{equation}
In the notation of Lemma 3.5 of Ref.~\cite{dupuis_2014}, we can express this as
\begin{equation}
  \int d\psi_H f_\psi^2 = \btr{E(M)M},
\end{equation}
where $M = \ketbra{\chi}^{\otimes 2}$ and $E(M)$ is defined to be
\begin{equation}
  E(M) = \int dU_H (U\otimes U)M(U^\dagger \otimes U^\dagger).
\end{equation}
According to Prop. 2.2 in Ref.~\cite{collins_integration_2006} and Lemma 3.5 in Ref.~\cite{dupuis_2014}, it follows from Schur-Weyl duality that
\begin{equation}
  E(M) = \alpha\mathbb{1} + \beta F,
\end{equation}
where $F$ is the swap operator and the coefficients $\alpha,\beta$ satisfy $\alpha D^2 + \beta D = \btr{M}$ and $\alpha D + \beta D^2 = \btr{MF}$.
Here we have $M = \ketbra{\chi}^{\otimes 2}$ so $\btr{M} = \btr{MF} = 1$ and it follows that $\alpha = \beta = \frac{1}{D(D+1)}$.
Consequently, $\btr{E(M) M} = \frac{2}{D(D+1)}$.

\begin{acknowledgments}
  A.K., L.J.S., H.M.K., and C.M.B. acknowledge support from the Professional Research Experience Program (PREP) operated jointly by NIST and the University of Colorado. The authors thank Victor Albert and Adam Brandt for helpful comments. This work includes contributions of the National Institute of
  Standards and Technology, which are not subject to U.S. copyright. 
\end{acknowledgments}

\bibliographystyle{quantum}
\bibliography{benchmarking_paper}

\end{document}